\documentclass[%
aps,prb,superscriptaddress,twocolumn,floatfix,10pt]{revtex4-2}
\usepackage[utf8]{inputenc}
\usepackage{amsmath,amssymb}
\usepackage{empheq}
\usepackage{float}
\usepackage{lipsum}
\usepackage{mathtools,cuted}
\usepackage{esvect}
\usepackage{multirow}
\usepackage{xcolor}
\usepackage{soul}
\usepackage[export]{adjustbox}
\usepackage{graphicx}
\usepackage{dcolumn}
\usepackage{bm}
\usepackage{hyperref}
\usepackage{physics}
\usepackage[mathlines]{lineno}
\usepackage[capitalise]{cleveref}
\usepackage{todonotes}
\usepackage{bbm}
\usepackage{orcidlink}
\usepackage{tikz}
\usepackage[english]{babel} 
\begin{document}
	
	\title{Topology of honeycomb nanoribbons revisited}
	
	\author{Z.F. Osseweijer}
	\thanks{These authors contributed equally to this work.}
	\affiliation{%
		Institute for Theoretical Physics, Utrecht University, 3584CC Utrecht, The Netherlands\\
	}%

	\author{L. Eek}
	\thanks{These authors contributed equally to this work.}
	\affiliation{%
		Institute for Theoretical Physics, Utrecht University, 3584CC Utrecht, The Netherlands\\
	}%
	\author{H. J. W. Zandvliet}
	\affiliation{%
		Physics of Interfaces and Nanomaterials, $\text{MESA}^{+}$ Institute, \\
		University of Twente, Drienerlolaan 5, 7522NB Enschede, The Netherlands 
	}
	\author{P. Bampoulis}
	\affiliation{%
		Physics of Interfaces and Nanomaterials, $\text{MESA}^{+}$ Institute, \\
		University of Twente, Drienerlolaan 5, 7522NB Enschede, The Netherlands 
	}
	\author{C. Morais Smith}
	\affiliation{%
		Institute for Theoretical Physics, Utrecht University, 3584CC Utrecht, The Netherlands\\
	}%
	
	\date{\today}
	
	\begin{abstract}
		We present a in-depth study of end states in honeycomb nanoribbons, focusing on the interplay between nanoribbon termination, chiral symmetry, and complex next-nearest-neighbor hopping in the framework of the Haldane model. Although previous work has identified zero-dimensional end states in such systems, this analysis is incomplete. Here, we systematically investigate zigzag and armchair nanoribbons of various widths, using the multiband Zak phase to characterize the topological properties of the occupied bands. We show that the Zak phase is quantized only for certain ribbon terminations, and we elucidate how this termination dependence governs the existence and robustness of end states. Furthermore, we explore the effect of varying the complex next-nearest-neighbor hopping phase, demonstrating the breakdown of chiral symmetry, the evolution of the bulk gap, and the resulting depinning of end-state energies. Finally, we place our findings in the context of previous studies and discuss connections to the Kane–Mele model, including the role of Rashba spin-orbit coupling. Our work provides a more detailed analysis of topological end states in nanoribbons described by the Haldane and Kane-Mele models and offers a framework for their characterization in related systems.
	\end{abstract}
	
	\maketitle{}

	\section{\label{sec:Intro}Introduction
	}
	A defining feature of topological insulators (TIs) is the presence of in-gap states localized at the material's boundaries.  These boundary states are resilient against external perturbations by virtue of the topological nature of the bulk band structure. By considering the spectral symmetries and the dimensionality of the bulk, it is possible to classify the topological nature of a non-interacting fermionic system according to the tenfold way \cite{Tenfoldway}, which is based on the presence or absence of time-reversal, particle–hole, and chiral symmetries. 
	Each symmetry class, when combined with the system’s spatial dimensionality, admits a specific topological invariant, such as a Chern number, winding number, or $\mathbb{Z}_2$ invariant, dictating the existence and type of boundary modes. 
	
	The first topological phase discovered was the quantum Hall effect ($\mathbb{Z}$-type) \cite{klitzingNewMethodHighAccuracy1980}. Transport measurements at very low temperatures reveal quantized transverse (Hall) conductivity, when a strong magnetic field is applied perpendicularly to a two-dimensional (2D) electron gas. In 1988, Haldane showed that the requirement of a strong magnetic field to observe this effect in 2D could be relaxed to a requirement of broken time-reversal symmetry \cite{haldane_model_1988}. The inclusion of a finite but net-zero flux per unit cell generates a complex next-nearest-neighbor (NNN) hopping, which breaks time-reversal symmetry, and leads to a $\mathbb{Z}$-type topologically non-trivial Chern-insulating phase \cite{haldane_model_1988}. The relevance of this work was confirmed in 2005, when Kane and Mele generalized it with the introduction of spin-orbit coupling (SOC) in graphene, leading to the quantum spin Hall effect \cite{kaneQuantumSpinHall2005}. In the absence of a Rashba SOC, the Kane-Mele model reduces to two copies of the Haldane model, one for each spin. The intrinsic SOC breaks time-reversal symmetry for a single spin, but together the two spin species preserve time-reversal symmetry. Furthermore, it was shown that the intrinsic SOC would give rise to non-trivial phases, quantified by a $\mathbb{Z}_2$ topological invariant \cite{kaneTopologicalOrderQuantum2005}. The inclusion of a sizable Rashba SOC term closes the bulk gap and destroys the topological phase \cite{kaneTopologicalOrderQuantum2005}.

	In addition to spectral symmetries, crystalline symmetries, such as inversion or mirror symmetry, may also provide topological protection, leading to the concept of topological crystalline insulators (TCIs) \cite{fu_topological_2011}. Although TCIs host boundary modes, analogously to TIs, their protection is generally weaker, since it is typically easier to break a crystalline symmetry than a spectral symmetry, such as time-reversal. Crystalline symmetries depend on the precise spatial arrangement of atoms, and can therefore be destroyed by lattice distortions, disorder, or surface reconstructions, whereas spectral symmetries are intrinsic to the Hamiltonian and remain intact under a broader range of perturbations. Furthermore, while the topological modes in TIs exist regardless of the boundary termination, this is not necessarily the case for TCIs.

	An interesting question that arises from the characterization of topological materials according to the tenfold way, is what happens to the topological properties in-between integer dimensions. One way to approach this question is to consider topological systems with a fractal geometry, with a non-integer dimension as realized in fractal geometries \cite{Canyellas2024BismuthFractal,osseweijer_haldane_2024, eek_fractality-induced_2025}. 
	A second route is to investigate the behavior of systems upon reducing their dimensionality, i.e., making a 3D material thinner and thinner \cite{MoesBi2Se3} or a 2D material narrower and narrower \cite{traverso_emerging_2024, Klaassen2025GermaneneTI, Eek2025Electric}. Ref.~\cite{traverso_emerging_2024} proposed that zigzag Haldane nanoribbons undergo a transition from a 2D Chern insulator to a 1D topological insulator below a critical width. This behavior can be understood by realizing that, for decreasing ribbon width, the chiral edge modes of the nanoribbons start to hybridize, resulting in a hybridization gap, which may host topological end states. The emergence of such end states was later experimentally observed in germanene nanoribbons \cite{Klaassen2025GermaneneTI, Eek2025Electric}. These are well-described by the Kane-Mele model \cite{kaneQuantumSpinHall2005} which, in the absence of Rashba SOC, reduces to two time-reversal symmetric copies of the Haldane model. 
	
	While these works capture the emergence of zero-dimensional end states in Haldane nanoribbons, the previous analysis is not complete, and a full understanding requires further investigation. In the present work, we provide an in-depth study of end states in Haldane nanoribbons, and reveal additional aspects overlooked so far, such as an even/odd effect in the nanoribbon width. Our results emphasize the importance of choosing the correct unit cell corresponding to the end termination of the nanoribbons when considering their topological properties.
	
	The outline of this paper is as follows; we introduce the Haldane model on a ribbon geometry in Sec.~\ref{sec:haldane}. In Sec.~\ref{sec:Topo}, we define the multiband Zak phase, compute its value for zigzag and armchair ribbons of various widths, and discuss the Zak phase quantization. In Sec.~\ref{sec:topoinfluence}, we investigate the dependence of the topological properties on the nanoribbon termination and the effect of varying the flux $\phi$. In Sec.~\ref{sec:literature}, we compare our results to the previous theories in the literature. Then, we discuss the Kane-Mele model, with particular emphasis on the Rashba SOC in Sec.~\ref{sec:KM}. In Sec.~\ref{sec:exp}, we demonstrate how our theoretical results account for the findings of Refs.~\cite{Klaassen2025GermaneneTI, Eek2025Electric} and provide a coherent interpretation of the experimental observations. Finally, we present our conclusions in Sec.~\ref{Sec: Conclusion}.

	\section{Haldane model on nanoribbons}\label{sec:haldane}
	The Haldane model is a tight-binding model of spinless fermions on a honeycomb lattice, subject to a magnetic flux that integrates to net-zero over the unit cell. The magnetic flux breaks time-reversal symmetry and drives the system into a topologically non-trivial insulating phase, such that protected chiral modes arise on its boundaries \cite{haldane_model_1988}. The corresponding Hamiltonian reads
	\begin{equation}
		H= t\sum_{\langle ij\rangle}c_i^\dagger c_j^{} + t_2\sum_{\langle\!\langle ij\rangle\!\rangle} e^{i\phi_{ij}}c_i^\dagger c_j^{}+\sum_i M_i c_i^\dagger c_i^{}.
		\label{eq:Hal}
	\end{equation}
	Here, the first term describes the nearest-neighbor (NN) hopping with amplitude $t$ and the second term describes the complex NNN hopping with amplitude $t_2$ and a phase $\phi_{ij}$, which encodes the magnetic flux $\phi$, such that  $\phi_{ij} = \phi$ ($-\phi$) for clockwise (counterclockwise) hopping. The last term captures the staggered mass $M_i = M$ ($-M$) for sites $i$ belonging to the A (B) sublattice \cite{haldane_model_1988}.
	
	When $M = t_2 = 0$, the Hamiltonian reduces to the well-known NN tight-binding model for graphene. Hence, the system is gapless and semi-metallic. The inclusion of either a finite $t_2$ or $M$ will open a gap, which is topological or trivial, respectively. The phase transition arising from the competition of these effects occurs at
	\begin{equation}
		M = \pm 3\sqrt3 t_2 \sin(\phi).
		\label{eq:phase_boundary}
	\end{equation}
	When $\abs{M} < \abs{3\sqrt3 t_2 \sin(\phi)}$, the system is classified as a $\mathbb{Z}$-type topological insulator. It hosts chiral edge modes and is characterized by a finite quantized Hall conductance \cite{haldane_model_1988}. Otherwise, the system is a trivial insulator.

	\begin{figure}
		\centering
		\includegraphics{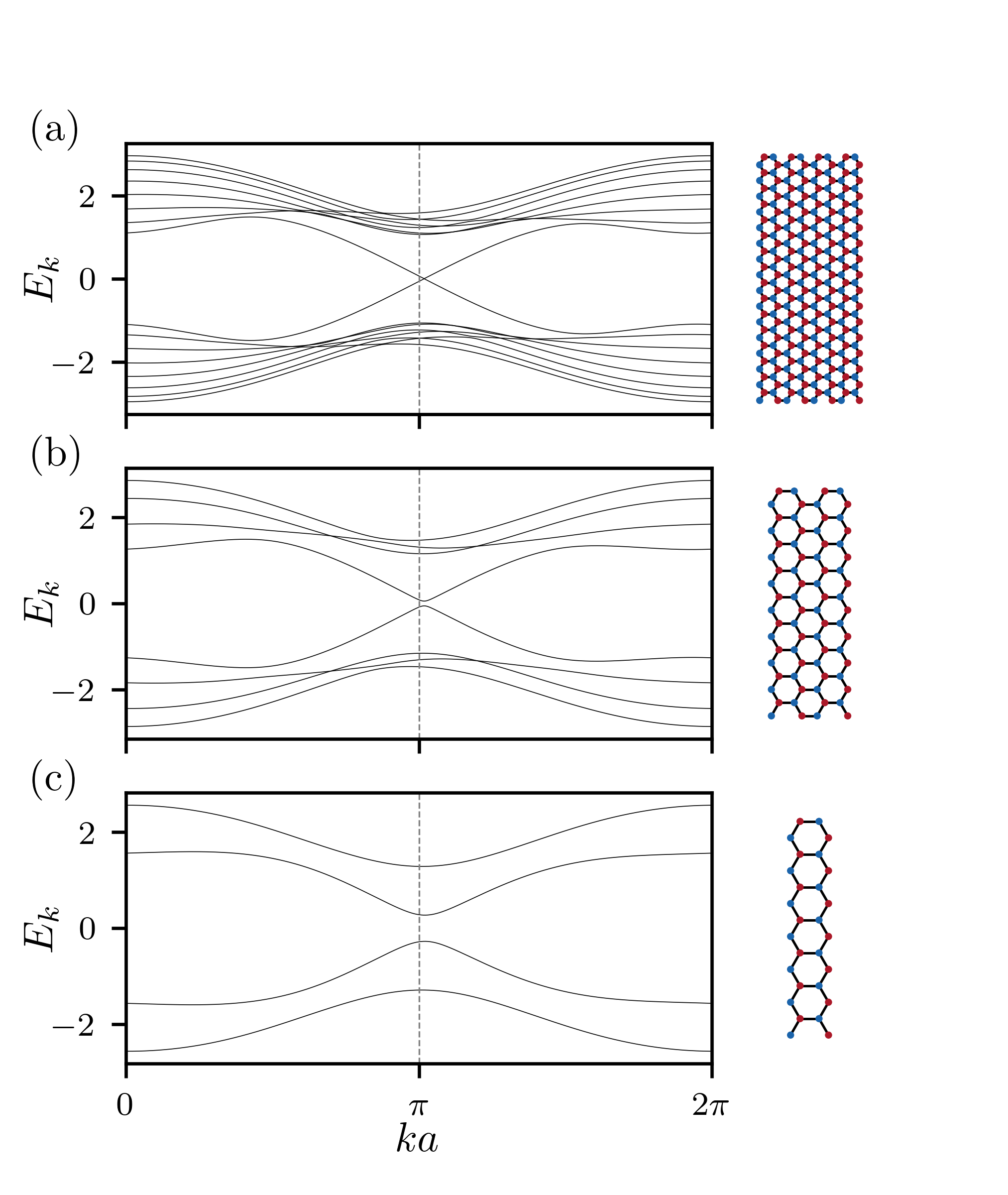}
		\caption{The bulk spectra of one-dimensional `zigzag' nanoribbons. The width of the considered nanoribbons is (a) 7 hexagons, (b) 3 hexagons, and (c) 1 hexagon. As the width decreases, a gap opens due to the real-space hybridization of the chiral edge modes. This hybridization gap increases in size as the width of the nanoribbon decreases. Here, $M = -0.5$, $t_2 = 0.3$, and $a$ is the lattice constant.}
		\label{fig:HybridizationGap}
	\end{figure}
	In this work, we investigate the Haldane model on nanoribbon geometries. To this end, we consider periodic boundary conditions in one direction and a finite width in the other, transverse, direction. The resulting system is a quasi one-dimensional (1D) crystal. Here, and in the remainder of this work, we set $t_2=0.3t$ and $\phi = \pi/2$, unless specified otherwise.
	
	In Fig.~\ref{fig:HybridizationGap}, the bulk spectra are depicted for different widths of `zigzag' nanoribbons in the 2D topological phase. As the width decreases, the 1D edge modes hybridize and a gap opens. The spectrum of a 7-hexagon-wide nanoribbon displays the 1D chiral modes of the Haldane model, see Fig.~\ref{fig:HybridizationGap}(a). Decreasing the width to 3 hexagons leads to a gap opening, as shown in Fig.~\ref{fig:HybridizationGap}(b). Upon decreasing further the width, until only one hexagon remains, we obtain the spectrum displayed in Fig.~\ref{fig:HybridizationGap}(c), where the size of the gap has increased.
	
	Since this gap is the result of the hybridization of topological edge modes, we will refer to it as the \textit{hybridization gap}. It is worth investigating whether any topological properties remain in the transition from a 2D system to 1D in the presence of the hybridization gap. 
	
	\section{Topological characterization}\label{sec:Topo}
	\subsection{Zak Phase}\label{sec:zak}
	To characterize the 1D topological nature of these systems, we will calculate the Zak phase $\varphi$ \cite{zak_berrys_1989}. For an isolated band, it reads
	\begin{equation}
		\varphi = \oint_{\text{BZ}} dk A(k) = i\oint_{\text{BZ}}dk\mel{u_k}{\partial_k}{u_k}.
	\end{equation}
	Here, the integral is performed over the 1D nanoribbon Brillouin zone (BZ); $k \in [0,2\pi/a)$. Furthermore, the Berry connection, $A(k)$, is given in terms of the occupied eigenvectors $\ket{u_k}$.  However, the unit cell of a nanoribbon consists of multiple bulk unit cells. Consequently, there are many bands, and they may cross, such that individual bands can no longer be treated independently. In this situation, the Zak phase is only well-defined for an isolated group of bands. For a given energy gap, all bands lying below it constitute an isolated set, and therefore admit a well-defined (multi-band) Zak phase. To compute this quantity, we employ the multiband generalization of the Zak phase, formulated in Refs.~\cite{soluyanov_smooth_2012, vanderbilt_berry_2018}.
	
	First, the 1D ribbon Brillouin zone is discretized into $M$ steps; the crystal momentum is $k_ja~=~2\pi j/(M-1)$, where $j \in \{0, 1, ..., M-1\}$, such that $k_0a = 0$ and $k_{M-1}a = 2\pi$. For each $k$-point, the $N\times N$ multiband Bloch Hamiltonian of a nanoribbon is diagonalized, resulting in a set of eigenstates
	\begin{equation*}
		\left \{ 
		U_{k_0}', U_{k_1}', \dots U_{k_{M-2}}', U_{k_{M-1}}'
		\right \},
	\end{equation*}
	where $U_{k_i}' = \{ |u_{k_i,0}'\rangle,  |u_{k_i,1}'\rangle, \dots  |u_{k_i,N-1}'\rangle\}$, i.e. colums of $U_{k_i}'$ are the Bloch vectors $|u_{k_i,n}'\rangle$. Of these, we only retain the isolated set of $\ell$ bands of interest, such that our set of states has dimension $M\times N \times \ell$. Before the Zak phase can be calculated, the $U(1)$ gauge degree of freedom in the eigenstates of the Hamiltonian has to be considered. When using numerical methods to diagonalize the Hamiltonian, the gauge is chosen at random. Therefore, to compare eigenvectors at different points in the Brillouin zone, the states must be transformed into the same choice of gauge. We will denote a non-gauge-fixed quantity by a prime (i.e. $|u'\rangle$), and the gauge-fixed quantities by the unprimed counterpart (i.e. $|u\rangle$). We choose a consistent gauge by transforming the set of filled bands to the parallel transport gauge, which is defined such that the overlap matrix between any two neighboring sets of states with elements $L_{mn} = \braket{u_{k_j,m}}{u_{k_j+1,n}}$ is Hermitian and has positive eigenvalues. To achieve this, we take the set of filled bands at $k=0$ to be in the correct gauge. Then, for each $k$-step, the overlap matrix $L'_{mn} = \langle u'_{k_j,m}| u'_{k_{j+1},n}\rangle$ is calculated and a singular value decomposition (SVD) $L'=V\Sigma W$ is evaluated. We then align the set of filled bands at $k_{j+1}$ by acting on them with the transformation $WV^\dag$, 
	\begin{equation}
		U_{k_{j+1}} = U_{k_{j+1}}' WV^\dagger \, .
	\end{equation}
	In this way, the gauge-fixed overlap $L$ becomes Hermitian, and its eigenvalues will be $\Sigma_{ii}$, which are positive by definition of the SVD. The crux is that for a topological phase, it is impossible to define a global smooth gauge by virtue of the topological obstruction. 
	By smoothing out the gauge, we have effectively accounted for the arbitrary numerical gauge and shifted the topological gauge discontinuity to the edge of the Brillouin zone. We obtain topological information by comparing the set of states at $k=0$ and $k=2\pi/a$. 
	With a consistent set of eigenstates, it only remains to calculate the Zak phase. 
	To do so, we need to compare the overlap matrix $S(k_0, k_{M-1})$ between the set of filled bands at $k=0$ and $k=2\pi/a$. Explicitly, the Zak phase corresponding to the system in which $n$ bands are filled is given by
	\begin{equation*}
		\varphi^{(n)} =  -\Im\log[\det S(k_0, k_{M-1})],
	\end{equation*}  
	where the matrix elements of $S(k_0, k_{M-1})$ are given by $S_{mn}(k_0, k_{M-1}) = \langle u_{k_0,m} | u_{k_{M-1},n}\rangle$. Here, we must ensure that this quantity is on the principal branch, as the Zak phase is only defined modulo $2\pi$. A quantized Zak phase is directly related to the appearance of topological end states, and these states remain pinned in energy throughout the topological phase \cite{vanderbilt_berry_2018}, a feature of significance in the remainder of this work. 
	
	After having established a method to calculate the Zak phase of a gap in a generic multiband setting, we turn our attention to the hybridization gap in narrow nanoribbons. 
	In general, there are infinitely many ways to cut a narrow nanoribbon from a 2D flake, parameterized by the angle of cutting \cite{term1, term2}. 
	In this work, we focus on two common choices: one corresponding to a `zigzag' termination and one to an `armchair' termination. In these cases, the termination direction is either parallel or orthogonal to the intracell bond, respectively. 
	
	\begin{figure*}
		\centering
		\includegraphics{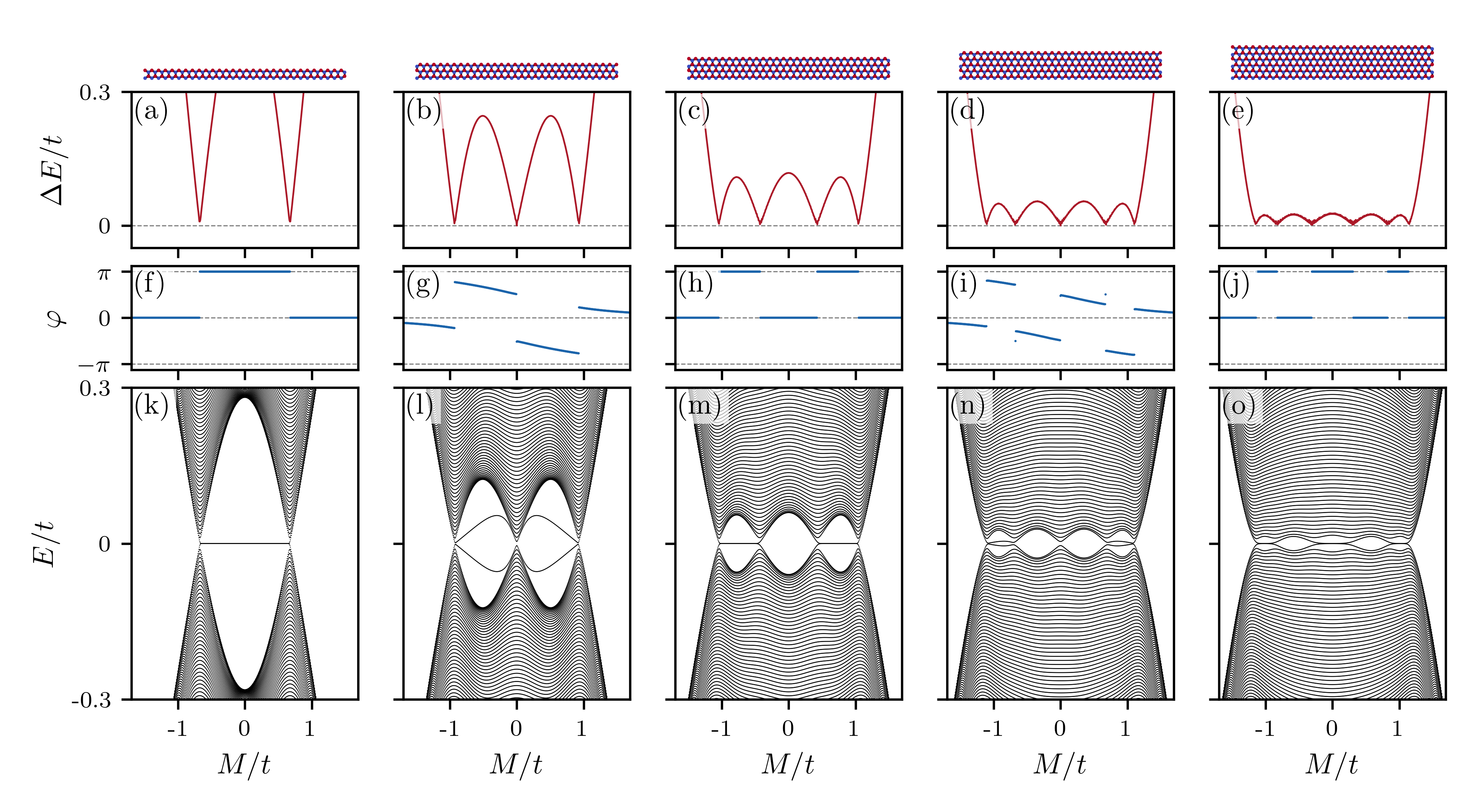}
		\caption{The energy gap, Zak phase, and OBC spectrum of zigzag Haldane nanoribbons as a function of $M$. (a)-(e), the bulk energy gap of the nanoribbons, (f)-(j), the Zak phase in the bulk gap, and (k)-(o) the OBC spectrum of a 500-unit-cell-long nanoribbon. These properties are shown for nanoribbons of different widths: (a, f, k) 1-hexagon, (b, g, l) 2-hexagon, (c, h, m) 3-hexagon, (d, i, n) 4-hexagon, and (e, j, o) 5-hexagon wide nanoribbons. Here, $t_2=0.3t$.}
		\label{fig:zigzag}
	\end{figure*}
	
	\begin{figure*}
		\centering
		\includegraphics{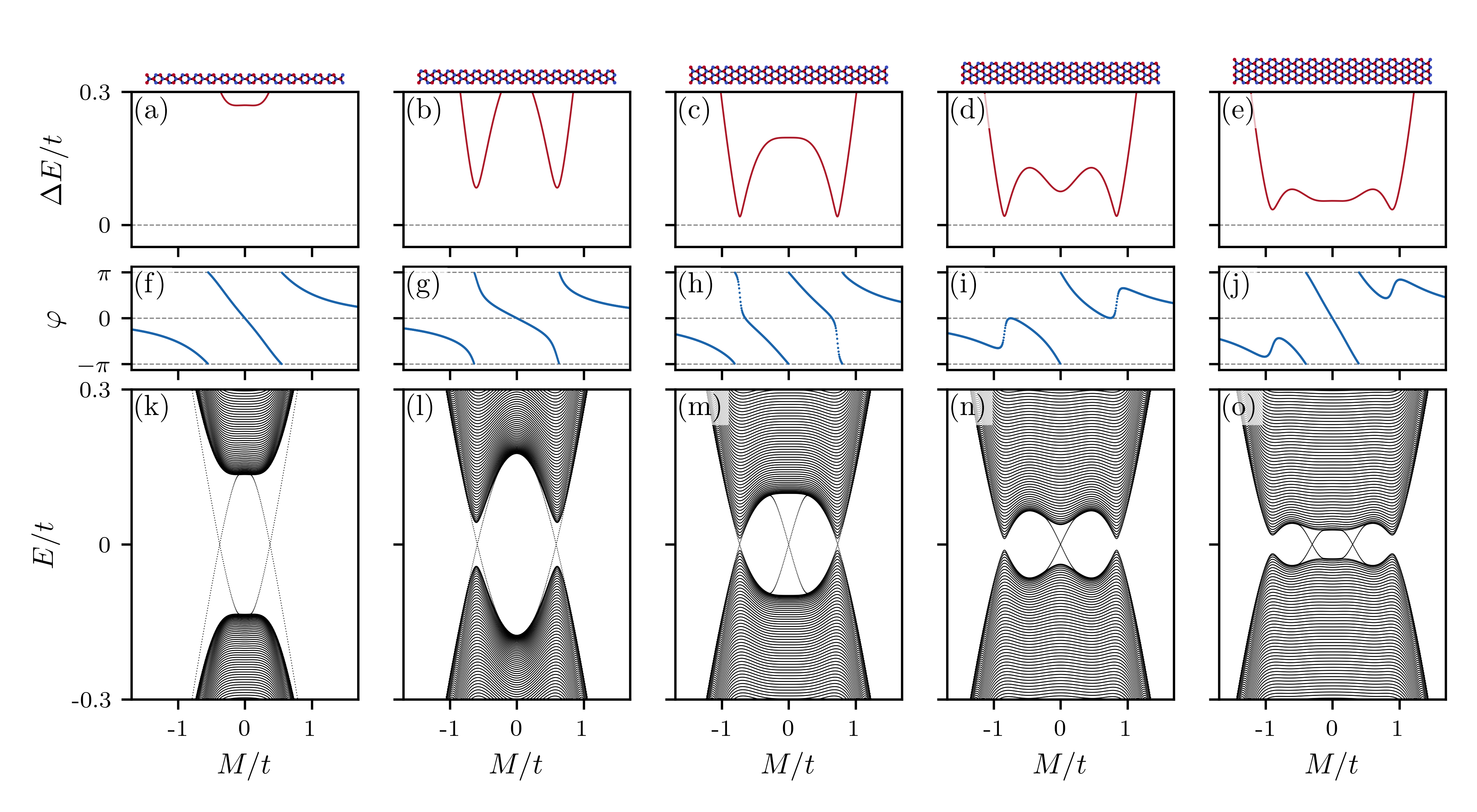}
		\caption{The energy gap, Zak phase, and OBC spectrum of armchair Haldane nanoribbons as a function of $M$. (a)-(e), the bulk energy gap of the nanoribbons, (f)-(j), the Zak phase in the corresponding gap, and (k)-(0) the OBC spectrum of a 500-unit-cell-long nanoribbon. Note that these are not dispersing modes, but the evolution of OBC spectra with $M$. These properties are shown for nanoribbons of different widths: (a, f, k) 1-hexagon, (b, g, l) 1.5-hexagon, (c, h, m) 2-hexagon, (d, i, n) 2.5-hexagon, and (e, j, o) 3-hexagon wide nanoribbons.}
		\label{fig:armchair}
	\end{figure*}
	
	\subsection{Zigzag nanoribbons}\label{sec:Topo_ribbon}
	In Fig.~\ref{fig:zigzag}, we present the bulk hybridization gap $\Delta E$, the multiband Zak phase $\varphi$, and the energy spectrum of a finite-size nanoribbon as a function of the staggered mass $M$. Each column corresponds to a nanoribbon of a different width. From left to right, we have 1-, 2-, 3-, 4-, and 5-hexagon-wide nanoribbons, as depicted in the upper row. Figs.~\ref{fig:zigzag}(a)-(e) reveal that an $n$-honeycomb-wide nanoribbon has $n+1$ gap closings, and that the size of the hybridization gap decreases as the width increases. 
	Figs.~\ref{fig:zigzag}(f)-(j) display the Zak phase evolution with $M$ \footnote{The spurious isolated dots in Fig.~\ref{fig:zigzag}(g), correspond to gapless realizations, for which the Zak phase is ill-defined and has no physical meaning.}. Importantly, the quantization of the Zak phase exhibits an alternating pattern, with the Zak phase only being quantized for nanoribbons of odd width. In even-width nanoribbons, quantization is lacking, but discontinuities appear whenever the gap closes. This is in contrast to Refs.~\cite{traverso_emerging_2024, Klaassen2025GermaneneTI, Eek2025Electric}, where such an even/odd effect is not reported. 
	This result is corroborated by the spectra for open boundary conditions (OBC) in Figs.~\ref{fig:zigzag}(k)-(o), where we observe $E=0$ pinned end states only in the odd-width nanoribbons. Although the end states are present in even-width nanoribbons, their energy is not pinned to zero because of the lack of topological protection, as predicted by the non-quantization of the Zak phase.
	
	\subsection{Armchair nanoribbons.} 
	Now, we present similar calculations for armchair nanoribbons. Here, the width of the nanoribbons can additionally increase by half a hexagon, while keeping a `smooth' edge, see e.g. the top row of Fig.~\ref{fig:armchair}. In this figure, the size of the bulk hybridization gap $\Delta E$, the multiband Zak phase $\varphi$, and the energy spectrum of a finite-size nanoribbon are shown as a function of the staggered mass $M$, analogous to Fig.~\ref{fig:zigzag}. For the armchair termination, we present results for 1-, 1.5-, 2-, 2.5- and 3-hexagon-wide nanoribbons. Figs.~\ref{fig:armchair}(a)-(e) indicate that for narrow nanoribbons the bulk gap no longer closes. Therefore, no phase transitions are expected in these systems. Indeed, Figs.~\ref{fig:armchair}(f)-(j) present no jumps in the Zak phase, and additionally, no quantization is observed. This is corroborated by the OBC spectra in Figs.~\ref{fig:armchair}(k)-(o), where no robustly pinned zero-energy end states appear. Instead, there can be in-gap states, which split as a function of $M$. Saliently, when these in-gap states cross, we observe $\varphi =\pi$. Similar behavior has been observed in topological pumping of aperiodic systems, which was explained by an emergent inversion symmetry \cite{Moustajpump}.
	
	\subsection{Even/odd effect}\label{sec:quantization}
	For odd widths, the Zak phase of the zigzag nanoribbons is quantized, whereas for even width it is not. Unlike topological invariants such as winding/Chern numbers, the Zak phase is not by definition quantized. Its quantization requires additional symmetries, such as a mirror or chiral symmetry. Since the Zak phase in Fig.~\ref{fig:zigzag}(f, h, j) is quantized, there should be a symmetry that is responsible for this behavior. Indeed, the Bloch Hamiltonian of these odd-width nanoribbons has a chiral symmetry,
	\begin{equation}
		\Gamma h(k) \Gamma^{-1} = -h(k), \label{eq:chiral}
	\end{equation}
	with
	\begin{equation}
		\Gamma = \sigma_X \otimes \sigma_y,
	\end{equation}
	where $\sigma_X$ is the $2n \times 2n$ matrix with ones on its anti-diagonal, and $\sigma_y$ is the Pauli $y$ matrix (here $4n$ is the amount of atoms in the unit cell). Consequently, the ribbon Hamiltonian may be written in the chiral basis through
	\begin{equation}
		\tilde{h}(k) = U h(k) U^\dagger = \begin{pmatrix}
			0 & q(k) \\ q^\dagger(k) & 0
		\end{pmatrix},
	\end{equation}
	where $q(k)$ is the matrix describing the coupling between the two chiral subspaces. The unitary matrix $U$ is obtained from diagonalizing $\Gamma$, i.e.
	\begin{equation}
		\tilde{\Gamma} = U\Gamma U^\dagger =\begin{pmatrix}
			\mathbb{I} & 0 \\ 0 & -\mathbb{I}
		\end{pmatrix},
	\end{equation}
	where $\mathbb{I}$ is the $2n\times 2n$ identity matrix. By virtue of the chiral symmetry, the topology of this system is captured by a winding number \cite{winding}
	\begin{equation}
		\nu = \frac{i}{2\pi}\int \text{d} k \text{Tr} \left[ q(k) \partial_k q^\dagger(k)\right].
	\end{equation}
	In chiral-symmetric systems, the Zak phase can be related to the winding number $\nu$ through $\varphi = \nu \pi \pmod{2\pi}$. This explains why the Zak phase is quantized only for odd-width zigzag nanoribbons. 
	\begin{figure*}
		\centering
		\includegraphics{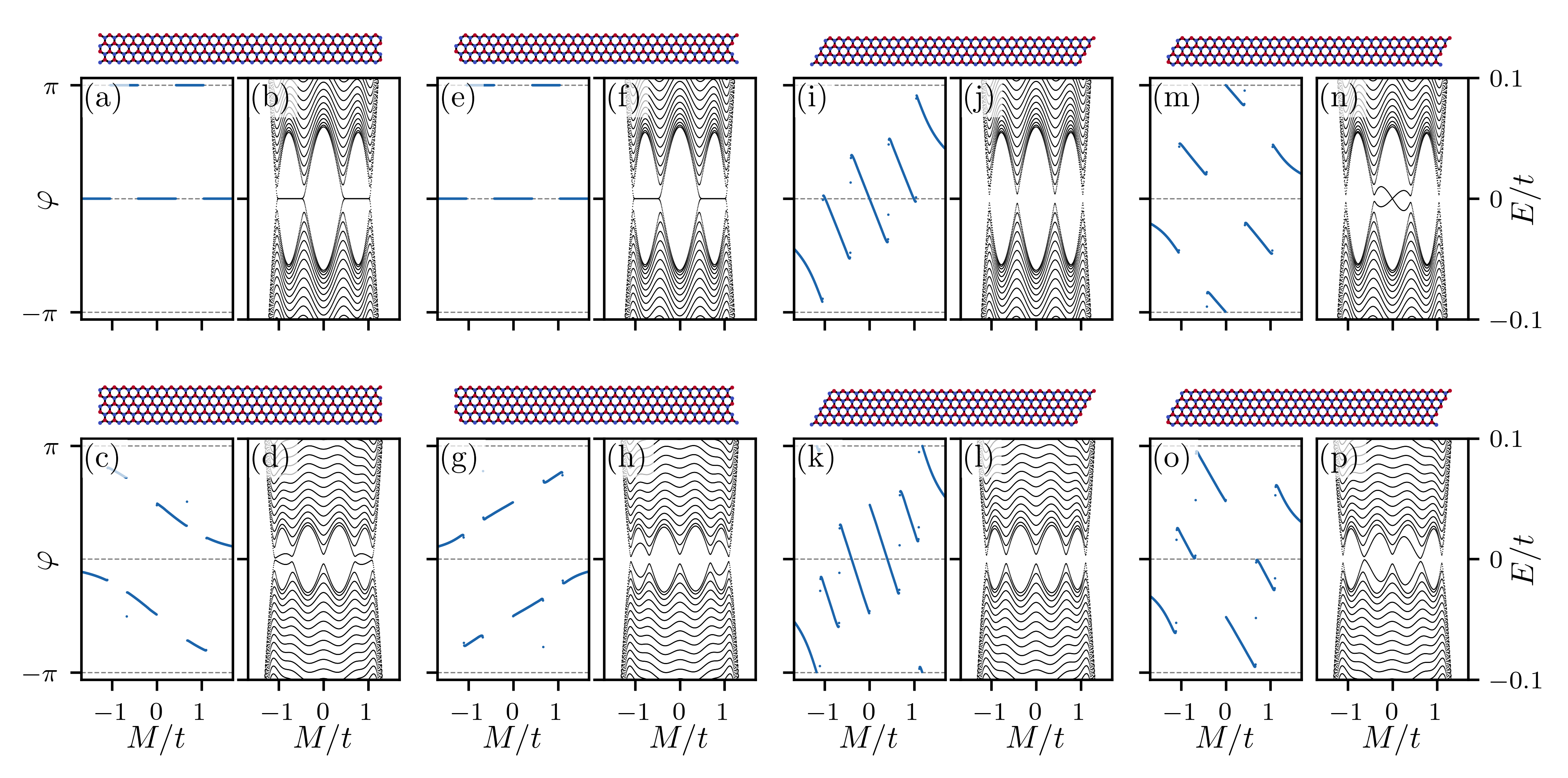}
		\caption{The Zak phase and open spectra corresponding to an odd (top) and even (bottom) zigzag nanoribbons with different terminations. In (a)-(d), the Zak phase [(a), (c)] and the spectrum of a 500-unit-cell-long nanoribbon [(b), (d)] are shown for a rectangular zigzag nanoribbon. Analogously, (e)-(h) show the results on a modified rectangular zigzag nanoribbon, (i)-(l) a rhombic nanoribbon, and (m)-(p) a modified rhombic nanoribbon. Quantization of the Zak phase is only observed in (a) and (e) because these are the only systems with a chiral symmetry.}
		\label{fig:termination}
	\end{figure*}
	Although chiral symmetry is often understood as a spectral symmetry, its emergence only for odd-width zigzag ribbons gives the emergent chiral symmetry a crystalline character, reminiscent of a topological crystalline insulator.

	\section{Factors influencing topological behavior}\label{sec:topoinfluence}
	\subsection{Dependence on end termination}\label{sec:Termination}
	Although the Bloch Hamiltonian encodes the bulk spectral properties of a system, it is still related to its termination. In an open geometry, a system must terminate in full copies of the unit cell if one is to use the bulk topological properties to discuss effects on the boundaries. Therefore, the termination of the systems intrinsically defines the unit cell. Although the eigenvalues are invariant under a redefinition of the unit cell, the eigenvectors and their properties, such as their topological character, can change, as known for the Su-Schrieffer-Heeger model \cite{su_solitons_1979}. Therefore, the termination dependence of the topological behavior in Haldane nanoribbons is captured by the choice of a bulk unit cell.
	
	Because there exist a variety of possible end terminations and corresponding unit cells, we now investigate a few examples of different end terminations. In Fig.~\ref{fig:termination}, we compare odd and even zigzag nanoribbons with four different terminations. We consider a rectangular, a modified rectangular, a rhombic, and a modified rhombic terminated nanoribbon. The real-space representations of these different terminations can be found at the top of Fig.~\ref{fig:termination}. The modified nanoribbons are related to their counterparts by flipping the outermost atoms (dangling bonds) on the corner from one end to the other. The Bloch Hamiltonian for the different terminations is presented in App.~\ref{app:hamils}. For each termination, 3- (top) and 4-hexagon-wide (bottom) nanoribbons are considered, and both the Zak phase and the OBC spectrum of a 500-unit-cell-long nanoribbon are reported. 
	
	\subsubsection{Rectangular \& modified rectangular end termination}
	In Figs.~\ref{fig:termination}(a) and (b), the results for an odd-width rectangular nanoribbon are displayed, and in Figs.~\ref{fig:termination}(c) and (d), this is done for an even-width rectangular nanoribbon. Here, we see the quantization of the Zak phase only for odd widths, as discussed in Sec.~\ref{sec:quantization}. These results are corroborated by the ones obtained for the modified rectangular nanoribbon in Figs.~\ref{fig:termination}(e)-(h). Indeed, when the system is in a topological phase, the end modes are pinned at zero energy, independently of the specific termination, as long as this termination respects the topology-protecting chiral symmetry. Because the repositioning of the `dangling' bonds does not break the chiral symmetry, the edge states remain topologically protected. This is in contrast with the even width case, where the rectangular nanoribbon lacks the required chiral symmetry [i.e. there does not exist a matrix $\Gamma$ satisfying Eq.~\eqref{eq:chiral}], and therefore so does the modified rectangular nanoribbon. Now, the observed end state energies of Figs.~\ref{fig:termination}(d) and (h) differ.
	
	\subsubsection{Rhombic \& modified rhombic end termination}
	For the rhombic nanoribbons, no quantization of the Zak phase occurs, as shown in Figs.~\ref{fig:termination}(i) and (k). Furthermore, no in-gap states appear in Figs.~\ref{fig:termination}(j) and (l), neither for the even- nor for the odd-width nanoribbons. Compared with the modified rhombic nanoribbon, we observe that the change in termination induces a change in the end states. For the modified rhombic nanoribbon, end states do appear in the middle gaps, see Figs.~\ref{fig:termination}(n) and (p). These states lack topological protection, as illustrated by their non-zero energy and the lack of Zak phase quantization, visible in Figs.~\ref{fig:termination}(m) and (o).
	The results in Fig.~\ref{fig:termination} also elucidate the nature of the chiral symmetry; all zigzag ribbons with chiral symmetry have an apparent mirror symmetry over the $x$-axis that is only present for zigzag ribbons of odd width. Both the $t_2$ and $M$ terms in Eq.~\eqref{eq:Hal} break this mirror symmetry, but together they preserve the emergent chiral symmetry.
	
	For completeness, the results analogous to Fig.~\ref{fig:armchair} for two extra terminations of an armchair nanoribbon are included in App.~\ref{app:TerminationArmchair}.

	Finally, let us briefly comment on the influence of the choice of termination. The unit cells of honeycomb nanoribbons consist of multiple atoms, and therefore allow for more freedom in the way the nanoribbons can be terminated. We observe the appearance of (i) pinned/topological end states, (ii) split end states, (iii) no end states for a given value of $M$.
	By combining different terminations, the end state energies can be engineered.
	For example, combining the terminations shown above Figs.~\ref{fig:termination}(a) and \ref{fig:termination}(i) would yield a nanoribbon which hosts a single pinned end state, localized on its left termination, also known as monomodes \cite{PlateroMonomode,PhysRevResearch.6.023140, zhao_topological_2018, parto_edge-mode_2018, henriques_topological_2020}. 
	This behavior is comparable to the SSH model, where the choice of termination also strongly influences the low-energy OBC spectrum. When terminating in a half unit cell, the SSH model will always have a single end mode, regardless of the parameter choice (given that the system is gapped) \cite{parto_edge-mode_2018, henriques_topological_2020}. This can be understood by realizing that for such a setup, the SSH model always terminates in a weak bond at one end of the chain, hence allowing for a single in-gap end state. Unlike the SSH model, where such a mode is always present, in the nanoribbon the end mode can be switched off by varying the value of $M$. 
	In Sec.~\ref{sec:literature}, we will further elaborate on this topic by choosing a ribbon end termination such that the ends are \textit{symmetrized}.

	\subsection{Phase dependence}\label{sec:phi}
\begin{figure}
		\centering
		\includegraphics[width=\linewidth]{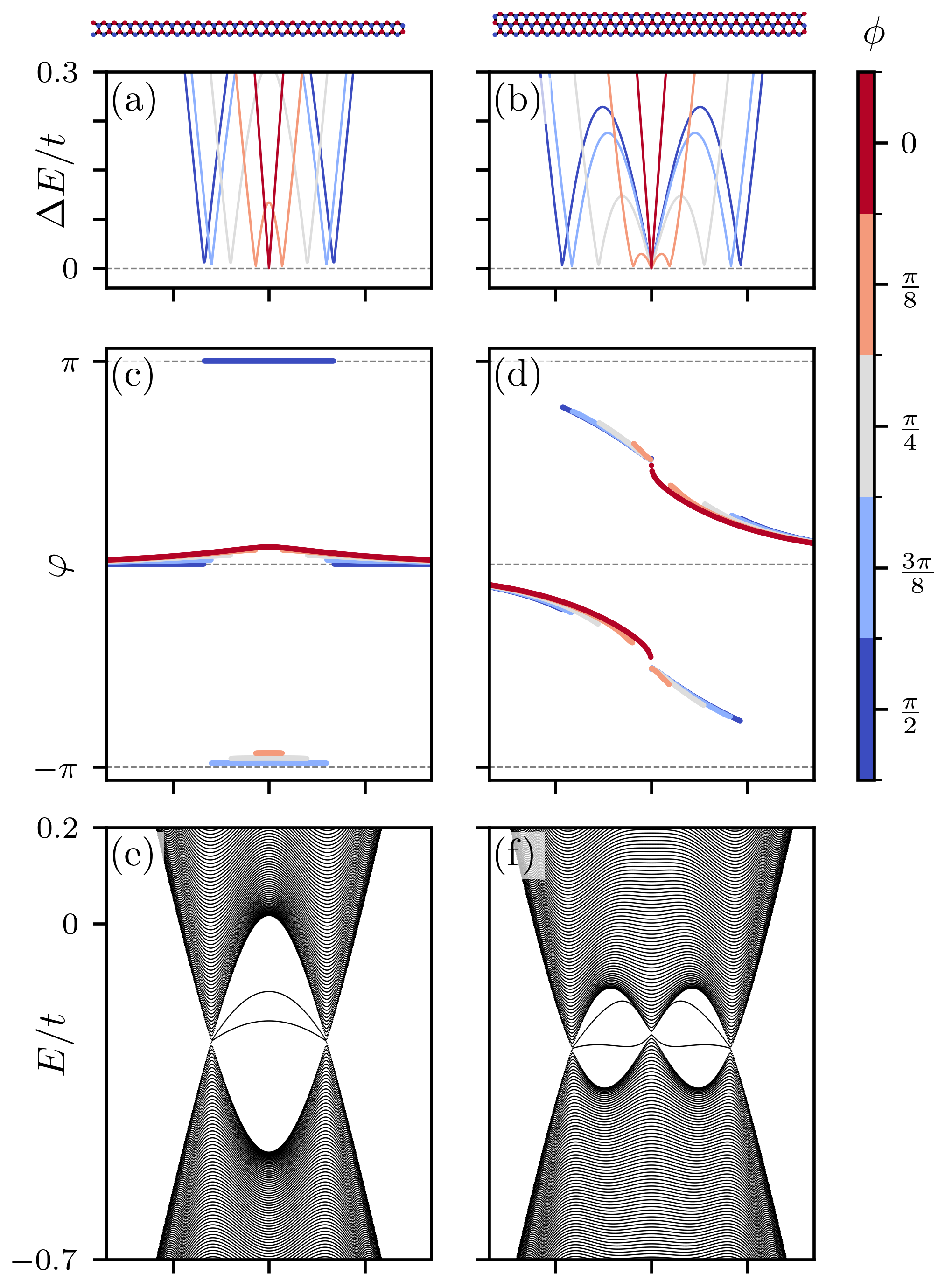}
		\caption{The complex NNN hopping phase $\phi$ dependence of the hybridization gap and the Zak phase as a function of staggered mass $M$ of odd [(a),(c), and (e)] and even [(b), (d), and (f)] zigzag nanoribbons, 1-hexagon- and 2-hexagons wide, respectively. In (a) and (b), the hybridization gap size is shown for 5 values of $\phi$, as indicated by the color bar. In both cases, the hybridization gap decreases in size as the complex hoppings are rotated onto the real axis, and vanishes when the NNN hopping has become fully real. Furthermore, the gap opening conditions become more constrained, and as the NNN hopping becomes real, the gap opens for smaller ranges of the staggered mass. In (c) and (d), we see that the chiral symmetry of the $\phi=\pi/2$ case is lost, because the Zak phase loses its quantization. Open spectra for $\phi=3\pi/8$ are shown in (e) and (f), where unpinned edge modes are observed. }
		\label{fig:phi}
	\end{figure}

	Now, we investigate how the Zak phase is affected when the NNN hopping acquires a real component, i.e. $\phi \neq\pi/2$. The chiral symmetry discussed in Sec.~\ref{sec:quantization} is preserved only when the NNN hopping is purely imaginary, corresponding to $\phi = \pm \pi/2$. In Fig.~\ref{fig:phi}, we show the dependence of both the bulk gap and the Zak phase on $\phi$ for a 1-hexagon- and 2-hexagon-wide zigzag nanoribbon. In Fig.~\ref{fig:phi}(a), one may observe how the critical value of the mass corresponding to the gap closing is reduced, and hence, how the size of the hybridization gap decreases as $\phi$ is varied from $\pi/2$ to $0$. For fully real NNN hoppings ($\phi=0$), the gap is closed and the system is semi-metallic. Similarly, in Fig.~\ref{fig:phi}(b), we see that the critical $M$ at which the gap closes decreases and the gap size becomes smaller. The Zak phase of these nanoribbons is shown in Figs.~\ref{fig:phi}(c) and \ref{fig:phi}(d). In Fig.~\ref{fig:phi}(c), one observes that the perfect quantization $\varphi=n\pi$ is lost for $\phi \neq \pi/2$. 
	Finally, Figs.~\ref{fig:phi}(e) and \ref{fig:phi}(f) present the OBC spectra for $\phi=3\pi/8$, showing the evolution of the end-state energies as a function of the staggered mass $M$. 
	The OBC spectra are no longer symmetric around $E=0$, which is a consequence of the broken particle-hole symmetry due to the finite real NNN hopping. Furthermore, the Zak phase is no longer quantized, and the end-state energies are no longer pinned to zero, consistent with the loss of chiral symmetry.
	
	\section{Comparison to literature}\label{sec:literature}
	\begin{figure}
		\centering
		\includegraphics{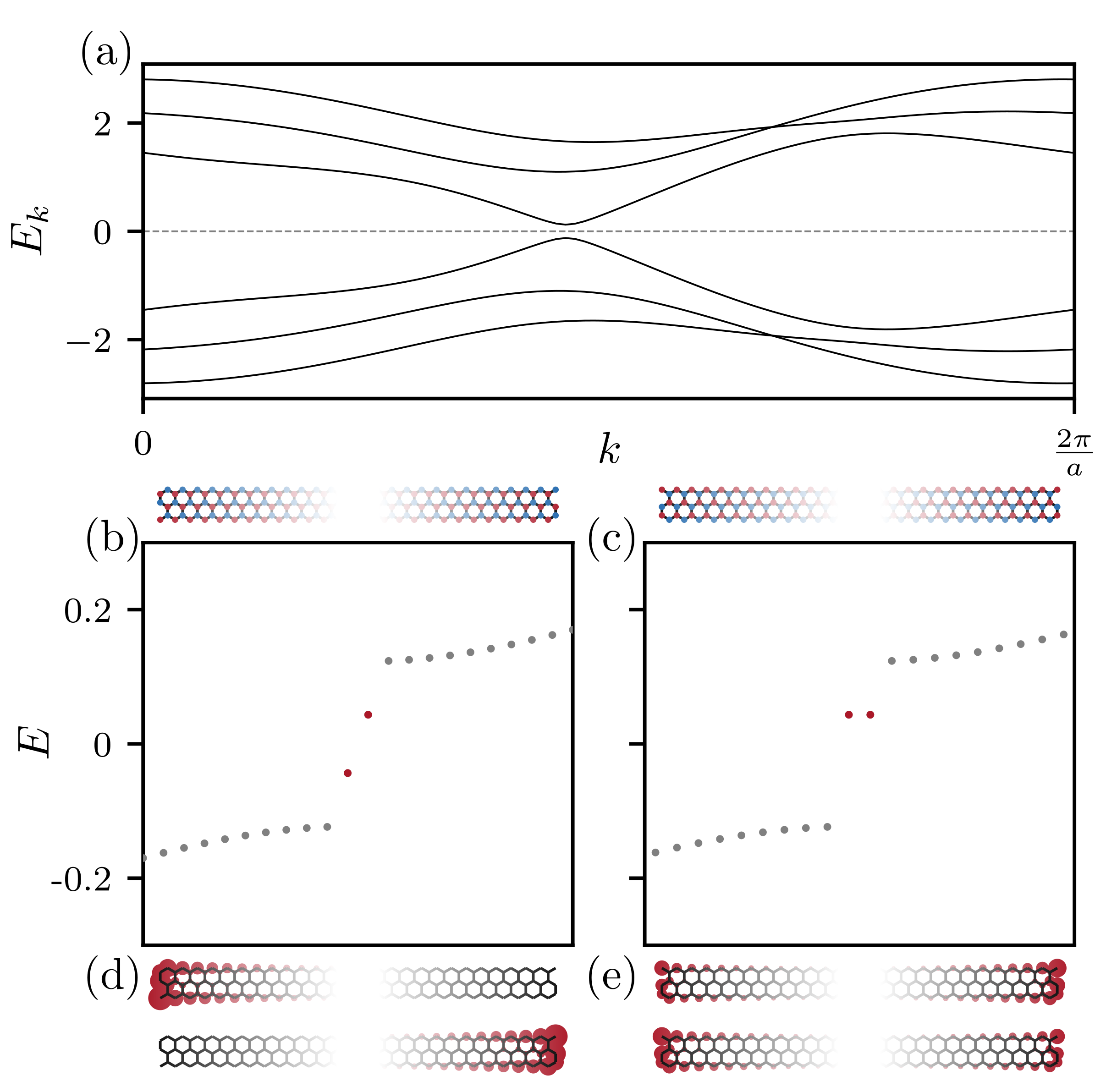}
		\caption{The effect of symmetrizing the boundaries of a 2-hexagon wide zigzag nanoribbon. (a) The bulk spectrum of the nanoribbon, (b) the low energy sector of the spectrum of a 500-unit-cell-long nanoribbon, (c) the low energy sector of the spectrum of a 499.5-unit-cell-long nanoribbon, which has symmetric boundaries. The energies of the end modes are marked in red, and in (d) and (e), the $\abs{\psi}^2$ of these end states is shown.}
		\label{fig:EndModeSplitting}
	\end{figure}
	The end states in Haldane zigzag nanoribbons were first reported by Ref. ~\cite{traverso_emerging_2024}. This was theoretically further investigated and experimentally realized in Refs.~\cite{Klaassen2025GermaneneTI, Eek2025Electric}. 
	
	However, in Ref. \cite{traverso_emerging_2024}, only a single Bloch Hamiltonian was presented, corresponding to a rhombic termination, while the open system was depicted with a rectangular termination. Therefore, the nanoribbon end terminations do not seem to match the Bloch Hamiltonian for which the authors calculate bulk topological invariants. As outlined in the previous sections, a bulk-boundary correspondence can only be invoked when the nanoribbon terminates in the same full unit cell used to calculate the Zak phase.
	
	Furthermore, the method proposed in Ref.~\cite{traverso_emerging_2024} to verify the topological character of the nanoribbons is based on the calculation of a quantity that in itself is not related uniquely to the physical Zak phase. 
	Instead, the supposed topological nature is claimed to be apparent after a comparison is performed to the same quantity in the atomic limit, i.e. $\varphi(M) - \lim_{M\rightarrow \pm\infty}\varphi(M)$. 
	Because the Zak phase is one-to-one related to physical observables such as the electric polarization \cite{vanderbilt_berry_2018}, this comparison step is unusual. Consequently, we believe that the calculation method is incorrect, and suspect the mistake to be in the position dependent unitary transformation $U(N)$ of the eigenstates. In the notation adopted in this work, this transformation would be $\ket{u_{k_{M-1}, n, l}} = \exp{(-2\pi ix_l/a)}\ket{u_{k_0, n, l}}$, where the exponential acts on the $l$-th component of the $n$-th band, and $x_l$ is the position along the periodic direction of the $l$-th basis element, i.e. atom. Such a transformation is necessary when changing tight-binding conventions, as described in Ref.~\cite{vanderbilt_berry_2018}. However, the system described in Ref.~\cite{traverso_emerging_2024} is already in the correct `unit-cell periodic' convention and does not require such a transformation.
	
	Additionally, Ref.~\cite{traverso_emerging_2024} presents OBC spectra and wavefunctions displaying pinned states for both odd- \textit{and} even-width nanoribbons, i.e., degenerate in-gap states are observed for the even-width nanoribbons, in contrast to the mass-induced splitting of end-state energies observed in the present work. 
	We hypothesize that the degeneracy obtained in Ref.~\cite{traverso_emerging_2024} originates from making both terminations symmetric by introducing a real-space inversion axis. 
	
	To achieve their results, one must terminate a nanoribbon not with a full copy of the bulk unit cell, but with `half' a unit cell \footnote{I.e. in a similar way to how we terminate the armchair nanoribbons in half hexagons in the open direction.}. 
	In Fig.~\ref{fig:EndModeSplitting}, we compare two nanoribbons with the same unit cell, but terminating with either a full unit cell (500 unit-cell long) or half a unit cell (499.5 unit-cell long). 
	In Fig.~\ref{fig:EndModeSplitting}(a), the band structure of the 1D bulk of a two-hexagon-wide nanoribbon is shown, and the finite-size hybridization gap is observed. As discussed in Sec.~\ref{sec:Topo}, although nanoribbons of this width host end modes, these end modes are not topologically protected. 
	
	Furthermore, in Figs.~\ref{fig:EndModeSplitting}(b) and (c), the low-energy sector of the spectrum of nanoribbons with full unit cell termination and half unit cell termination is depicted, respectively. The in-gap energies corresponding to the end states are marked in red. In Fig.~\ref{fig:EndModeSplitting}(b), we observe the splitting of energies, as expected from Sec.~\ref{sec:Topo}, while in Fig.~\ref{fig:EndModeSplitting}(c), the predicted degeneracy of end modes is observed. 
	In both cases, the energy is non-zero because these states localize strongly on the outermost atoms, which have a local imbalance in sublattice sites and therefore in staggered mass in the case of even width nanoribbons. When terminating in full unit cells, this local imbalance is compensated by the other end, which results in unpinned end modes at different energies. On the other hand, when terminating in a half-unit cell, both ends have the same local environment and consequently the end states will have the same energy and appear pinned. This is the reason that the end states presented in  Ref.~\cite{traverso_emerging_2024} appear pinned at non-zero energy, which only occurs for even-width nanoribbons.
	
	Lastly, in Figs.~\ref{fig:EndModeSplitting}(d) and (e), the wavefunction distribution $\abs{\psi}^2$ of states corresponding to the marked energies in Figs.~\ref{fig:EndModeSplitting}(b) and (c) are shown, respectively. Indeed, we observe in Fig.~\ref{fig:EndModeSplitting}(d) that the states are localized on either end of the nanoribbon and rapidly decay into the bulk. In contrast, the states in Fig.~\ref{fig:EndModeSplitting}(e) are nearly degenerate, which causes mixing and a non-vanishing wavefunction distribution on both ends for a single state.
	As there is no possible choice of unit cell that results in the symmetric terminations as in Figs.~\ref{fig:EndModeSplitting}(c) and (e), one cannot extract information on the behavior at the boundaries from a bulk model. This shows once more the importance of the choice of termination for the topological properties of these nanoribbons, as discussed in Sec.~\ref{sec:Termination}.

	\section{Kane-Mele Model on nanoribbons}\label{sec:KM}
	\begin{figure}
		\centering
		\includegraphics{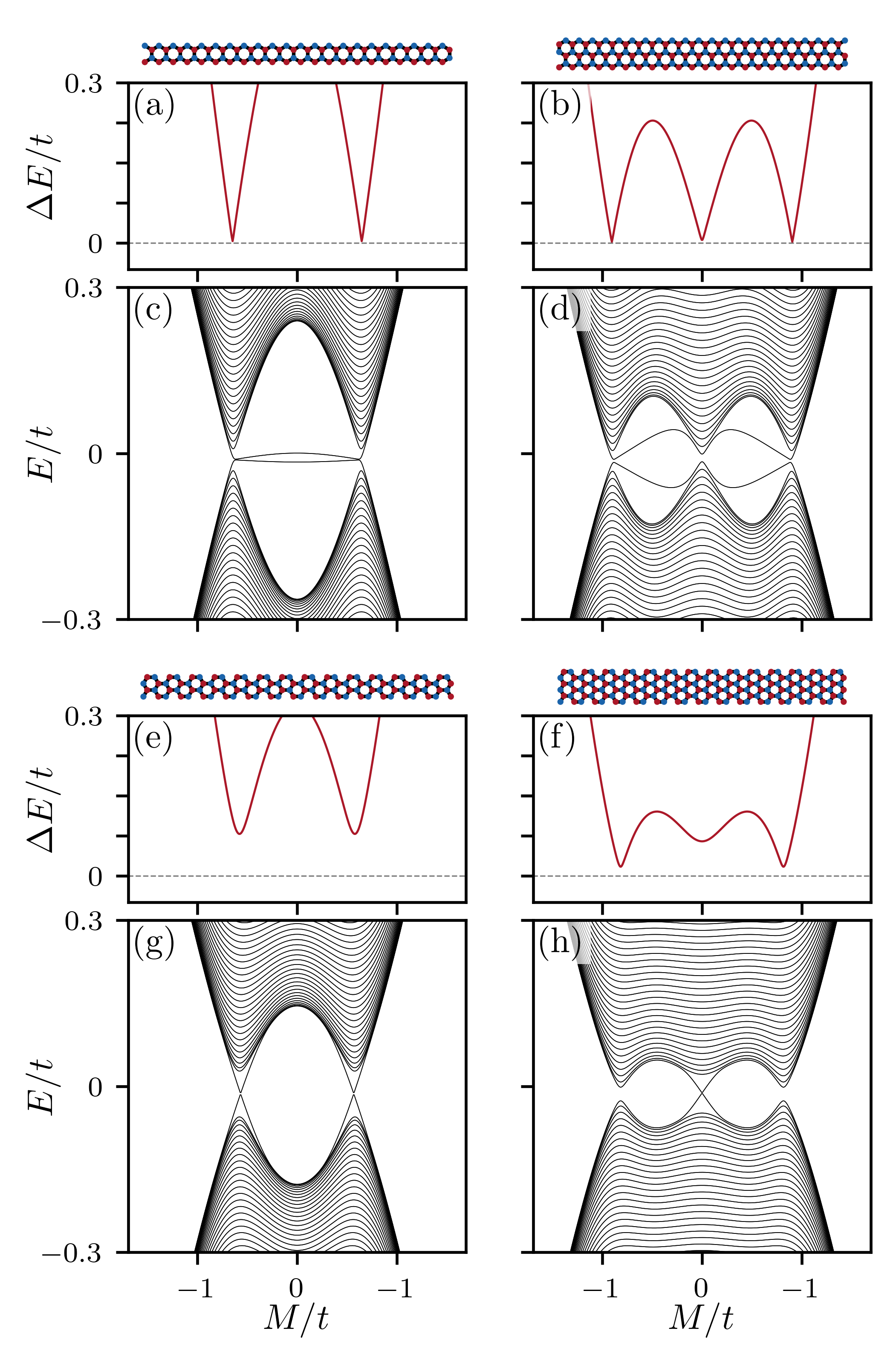}
		\caption{The effect of a finite Rashba SOC on the evolution of the spectra. In (a)-(d), zigzag nanoribbons of odd and even widths are considered. (a) and (b) depict the bulk spectrum, and (c) and (d) depict the spectrum of a 500-unit-cell-long nanoribbon. In (c), pinning is no longer observed as expected by the breaking of chiral symmetry caused by the Rashba SOC. In (e)-(h), similar results are presented for armchair nanoribbons, but here the effect of Rashba SOC is only observable in the broken particle-hole symmetry.}
		\label{fig:KaneMele}
	\end{figure}
	Finally, we investigate the Kane-Mele model for SOC in honeycomb materials \cite{kaneQuantumSpinHall2005,kaneTopologicalOrderQuantum2005}. The Kane-Mele model consists of two copies of the Haldane model, which map into each other under time-reversal symmetry. Each Haldane model describes a spin species with $t_2 = \pm \lambda_{SO}$. Additionally, the two spin species are coupled by a Rashba SOC. The full Kane-Mele Hamiltonian is given by \cite{kaneQuantumSpinHall2005, kaneTopologicalOrderQuantum2005}
	\begin{align*}
		H_{KM} &= t\sum_{\langle ij\rangle}\mathbf{c}_i^\dagger \sigma_0 \mathbf{c}_j^{} + \sum_i M_i \mathbf{c}_i^\dagger \sigma_0 \mathbf{c}_i^{} \\
		&+ i\lambda_{SO}\sum_{\langle\!\langle ij\rangle\!\rangle} \nu_{ij}\mathbf{c}_i^\dagger \sigma_z \mathbf{c}_j +i\lambda_R\sum_{\langle ij\rangle}\mathbf{c}_i^\dagger (\bm{\sigma} \times\bm{\hat{d}_{ij}}) \mathbf{c}_j^{}.
	\end{align*}
	Here, $\mathbf{c}_i = (c_{i,\uparrow}, c_{i,\downarrow})^T$, $\sigma_i$ are the Pauli matrices on spin space, $\bm{\sigma}$ the Pauli vector, $\bm{\hat{d}_{ij}}$ the unit vector pointing from site $i$ to $j$, and $i\nu_{ij} = \pm i$, analogous to $e^{i\phi_{ij}}$ in Eq.~\eqref{eq:Hal} for $\phi=\pi/2$.
	In the absence of Rashba SOC, the Kane-Mele model corresponds to two decoupled Haldane models, and therefore spin-degenerate pinned end states are expected to exist in the gap of an odd-width nanoribbon. Upon including Rashba SOC, Ref.~\cite{Klaassen2025GermaneneTI} argues that the effect on the end states is minimal, as Rashba SOC is a momentum term, i.e. $H_R \sim \alpha_R (k_x\sigma_y-k_y\sigma_x)$. Since the end states are zero-dimensional, they should not be too strongly influenced by the Rashba SOC. In Fig.~\ref{fig:KaneMele}, we present the hybridization gap size and the open spectrum of a Kane-Mele nanoribbon, with $\lambda_{SO}=0.3t$, and $\lambda_R=0.1t$. In Figs.~\ref{fig:KaneMele}(a)-(d), a zigzag nanoribbon is considered, analogous to the two left columns of Fig.~\ref{fig:zigzag}, but for the Kane-Mele model. Indeed, the results are similar. The most notable difference is in Fig.~\ref{fig:KaneMele}(c), where the previously pinned end states acquire a small energy splitting, indicative of broken chiral symmetry, caused by the inclusion of a Rashba term. Similarly, Figs.~\ref{fig:KaneMele}(e)-(h) are analogous to the left columns of Fig.~\ref{fig:armchair} and depict the effect of a Rashba term on the armchair nanoribbons. Here, the results are nearly identical, characteristic of the limited effect of the Rashba SOC on localized states.
	
	\section{Comparison with experiments}\label{sec:exp}
	Refs.~\cite{Eek2025Electric, Klaassen2025GermaneneTI} present experimentally realized end states on germanene nanoribbons. The even/odd effect discussed presently was previously overlooked, therefore it could be interesting to compare this new understanding with the experiments done in Refs.~\cite{Eek2025Electric, Klaassen2025GermaneneTI}. 
	
	The more stringent conditions for the emergence of topological states predicted here highlight an apparent mismatch between the theory and the experiments. Experimentally, the 2-hexagon-wide nanoribbons are consistent with end states near the middle of the bulk gap, whereas the 3-hexagon-wide ribbons show end states that are asymmetric with respect to the gap center. Within experimental uncertainty, this trend is opposite to that predicted by the present theory. However, a direct comparison of the theoretical results presented here with the scanning tunneling microscopy and spectroscopy experiments in Refs.~\cite{Eek2025Electric, Klaassen2025GermaneneTI} for germanene nanoribbons is not straightforward, because several important aspects of the experimental system are not included in the present theoretical calculations. 
	
	Firstly, the theoretical analysis is performed for free-standing nanoribbons, whereas the experimentally realized nanoribbons are supported on a substrate. This substrate not only breaks the inversion symmetry but may also introduce hybridization between the substrate and nanoribbon states. 
	In particular, the experimental germanene nanoribbons grown on Pt/Ge(110) studied in Refs.~\cite{Eek2025Electric, Klaassen2025GermaneneTI} exhibit a doubled periodicity. The origin of this doubled periodicity is still an open question, but could be related to charge density waves, the formation of a Luttinger liquid or reconstruction effects.
	Regardless of its origin, the experimental behavior is likely no longer fully described by a simple honeycomb lattice model. Consequently, its classification is more nuanced and may differ from the one proposed here.
	
	Furthermore, we note that the local electrostatic environment is not uniform along the nanoribbons in the experiments. The presence of charged defects or impurities in the substrate leads to charge puddles in the nanoribbons and consequently to a spatially varying shift of the energy states. As a result, the end-state energies and the local gap center are determined under different local conditions, which, along with electrostatic gating from the tip-induced electric field, complicates a quantitative comparison even further.
	
	To resolve this mismatch and perform a quantitatively reliable connection to experiment requires extending the model to include substrate effects and the experimentally observed doubled periodicity, possibly via ab initio density functional theory calculations. Nevertheless, the present results clearly demonstrate that ideal zigzag germanene and other honeycomb nanoribbons can host topological end states, with their existence sensitively depending on ribbon width, termination, and the associated unit-cell choice. This provides a useful starting point for understanding how such topological boundary states emerge and evolve in more realistic experimental settings.

	\section{Conclusions} \label{Sec: Conclusion}
	In this work, we present an in-depth analysis of (topological) end sates in Haldane nanoribbons, highlighting the crucial role of nanoribbon termination, and thus the choice of the bulk unit cell in determining their existence. By employing the multiband Zak phase, we demonstrate that topological quantization only occurs for odd-width zigzag nanoribbons with specific terminations, which give rise to an emergent chiral symmetry. Interestingly, this puts the system in an unusual position: The topology is protected by a spectral chiral symmetry, of which the existence is governed by a crystalline property, namely the termination, reminiscent of a crystalline topological insulator. We discuss how different terminations influence the chiral symmetry and, therefore, the end state spectrum. It is also briefly mentioned how this could be leveraged to induce monomodes; single end states.
	
	In addition, we investigate the effect of varying the flux phase $\phi$ and show that deviations from $\phi=\pi/2$ break the emergent chiral symmetry, lifting the pinning of the end-state energies away from zero energy. We also demonstrate that end states arise in the Kane–Mele model and that the inclusion of a finite Rashba SOC only induces a small splitting in the end-state spectrum.
	
	Moreover, we discuss the work by Ref.~\cite{traverso_emerging_2024}, on which the calculations in Refs.~\cite{Klaassen2025GermaneneTI, Eek2025Electric} are based. In these studies, the Zak phase is always quantized and the end-state energies appear to be pinned. 
	We explain the latter as a consequence of symmetrizing the end terminations, which enforces pinning albeit at non-zero energy, whereas the former arises from the mismatch between the bulk unit cell and the ribbon termination, compounded by an unnecessary unitary transformation. 
	We further believe that the theoretical method employed in Refs.~\cite{traverso_emerging_2024, Klaassen2025GermaneneTI} and \cite{Eek2025Electric} is incorrect, as it requires a comparison of the nanoribbon to its atomic limit when quantifying its topological character. 
	This is conceptually problematic because the Zak phase should directly correspond to the electric polarization, i.e. to a physical observable \cite{vanderbilt_berry_2018}. In contrast, the method presented here directly predicts the emergence of zero-energy topological end modes.
	
	Our results suggest several compelling directions for future work. One is to look for experiments on similar termination-dependent end states in other Chern insulating or quantum spin Hall materials, such as stanene, bismuthene, or WTe$_2$ \cite{LeeModified, LeeFloquet, stanene, bismuthene, Tang2017WTe2QSH, FinoHybrid}. It would be interesting to see whether the interplay between termination and symmetry that we identified here also shows up in those systems, or whether new types of boundary behavior emerge because these systems have different crystalline symmetries and orbital structure. Another promising direction is the dimensional crossover from 3D to 2D. Many 3D topological insulators, for example, Bi$_2$Se$_3$ or Bi$_2$Te$_3$, when sliced down to a few layers or patterned into narrow ribbons, develop edge or hinge states, whose properties depend sensitively on how the material is cut \cite{LiuOscillatory,Zhang2010Bi2Se3Crossover,MoesBi2Se3,TuningHOTI,Cutting,Schindler}. Understanding how these boundary modes evolve under confinement, and how much of the 3D topology survives in the 2D limit, could help to connect the physics of nanoribbons to a broader class of topological materials.

	\begin{acknowledgments}
		L.E. and C.M.S. acknowledge the research program “Materials for the Quantum Age” (QuMat) for financial support. This program (registration number 024.005.006) is part of the Gravitation program financed by the Dutch Ministry of Education, Culture and Science (OCW).
	\end{acknowledgments}

	\appendix
	
	\begin{widetext}
		\section{Bloch Hamiltonians of zigzag nanoribbons} \label{app:hamils}
		In the main text, we discuss how the choice of termination dictates the explicit form of the unit cell and its Bloch Hamiltonian. Here, we provide the Bloch Hamiltonians $H^{(i)}_z$ of the four-different-termination zigzag nanoribbons considered in the main text and in Fig.~\ref{fig:termination}. We define two helper functions, $g(k)$ and $f(k)$, which will greatly simplify the expressions. These are given by
		\begin{align*}
			g(\phi, k) &= t_2(e^{-i\phi}e^{ik}+e^{i\phi}e^{-ik}), & \text{and}& & f(\phi, k) &= t_2(e^{-i\phi} + e^{i\phi}e^{ik}).
		\end{align*}
		
		Using these definitions, the Bloch Hamiltonian of the rectangular nanoribbons considered in Fig.~\ref{fig:zigzag}, and Figs.~\ref{fig:termination}(a)-(d) becomes
		
		\usetikzlibrary{matrix,fit,backgrounds} 
		\begin{equation}
			\begin{tikzpicture}
				\node (label) at (-4,0) {$H^{(1)}_z(k) =$};
				\small
				\begin{scope}[shift={(4.1,0)}]
					\matrix (magic) [matrix of nodes,
					nodes={anchor=center},
					left delimiter={(},
					right delimiter={).}] {
						
						$M + g(\phi, k)$        & $t(1+e^{ik})$       & $f(\phi, k)$        & $0$                    & $0$                      & $0$                       & $\dots$ \\
						$t(1+e^{-ik})$          & $-M + g(-\phi, k)$  & $t$                 & $f(\phi, -k)$          & $0$                      & $0$                       & $\dots$ \\
						$f(-\phi, -k)$          & $t$                 & $M + g(\phi, k)$    & $t(1+e^{-ik})$         & $f(-\phi, -k)$           & $0$                       & $\dots$ \\
						$0$                     & $f(-\phi, k)$       & $t(1+e^{ik})$       & $-M + g(-\phi, k)$     & $t$                       & $f(-\phi, k)$             & $\dots$ \\
						$0$                     & $0$                 & $f(\phi, k)$        & $t$                    & $M + g(\phi, k)$         & $t(1+e^{ik})$             & $\dots$ \\
						$0$                     & $0$                 & $0$                 & $f(\phi, -k)$          & $t(1+e^{-ik})$           & $-M + g(-\phi, k)$        & $\dots$ \\
						$\vdots$                & $\vdots$            & $\vdots$            & $\vdots$               & $\vdots$                 & $\vdots$                  & $\ddots$ \\
					};
					\begin{scope}[on background layer={color=yellow}]
						\fill[blue!20] (magic-1-1.north west) rectangle (magic-3-3.south east);
					\end{scope}
					
				\end{scope}
			\end{tikzpicture}
			\label{eq:app_hamil_1}
		\end{equation}
		It consists of five distinct diagonals, which are either two- or four-periodic. An arbitrary-width ribbon can be obtained by repeating this pattern up to the desired width. The modified rectangular termination, considered in Figs.~\ref{fig:termination}(e-h), is made from $H^{(1)}_z$ by moving the `dangling bonds' to the right-hand side. This is achieved by taking $H^{(1)}_z(k)$ and changing the positions and connections of the first and last atoms, so that they are on the other side. Its Hamiltonian is therefore identical to $H^{(1)}_z(k)$, except for the first (marked blue in Eq.\ref{eq:app_hamil_1}) and last $3\times3$ block, which are given by
		\begin{align}
			H^{(2)}_z(k)_{1:3, 1:3} &=
			\begin{pmatrix}
				M + g(\phi, k) & t(1+e^{-ik}) & f(-\phi, -k) & \\
				t(1+e^{ik}) & -M +g(-\phi, k) & t  \\
				f(\phi, k) & t & M + g(\phi, k)
			\end{pmatrix} \text{, and} \nonumber \\
			H^{(2)}_z(k)_{n-2:n, n-2:n} &=     \begin{pmatrix}
				-M + g(-\phi, k) & t(1+e^{ik}) & f(-\phi, k) & \\
				t(1+e^{-ik}) & M +g(\phi, k) & t  \\
				f(\phi, -k) & t & -M + g(-\phi, k)
			\end{pmatrix}.
		\end{align}
		Similarly, for the rhombic nanoribbon of Figs.~\ref{fig:termination}(i)-(l) the Bloch Hamiltonian $H^{(3)}_z$ is given by,
		\begin{equation}
			\begin{tikzpicture}
				\node (label) at (-4,0) { $H^{(3)}_z(k) =$};
				\begin{scope}[shift={(4.1,0)}]
					\small
					\matrix (magic2) [matrix of nodes,
					nodes={anchor=center},
					left delimiter={(},
					right delimiter={), }] {
						$M + g(\phi, k)$ & $t(1+e^{ik})$ & $f(\phi, k)$ & $0$ & $0$ & $0$ & $\dots$\\
						$t(1+e^{-ik})$ & $-M +g(-\phi, k)$ & $t$ & $f(-\phi, k)$ & $0$  & $0$  & $\dots$\\
						$f(-\phi, -k)$ & $t$ & $M + g(\phi, k)$ & $t(1+e^{ik})$ & $f(\phi, k)$ & $0$ & $\dots$\\
						$0$ & $f(\phi, -k)$ & $t(1+e^{-ik})$ & $-M +g(-\phi, k)$ & $t$ & $f(-\phi, k)$ &$\dots$\\
						$0$ & $0$ & $f(-\phi, -k)$ & $t$ & $M+g(\phi, k)$ & $t(1+e^{ik})$ & $\dots$\\
						$0$ & $0$ & $0$ &$ f(\phi, -k)$ & $t(1+e^{-ik})$ & $-M+g(-\phi, k)$ & $\dots$ \\
						$\vdots$ & $\vdots$ & $\vdots$ & $\vdots$ & $\vdots$ & $\vdots$ & $\ddots$ \\
					};
					\begin{scope}[on background layer]
						\fill[blue!20] (magic2-1-1.north west) rectangle (magic2-3-3.south east);
					\end{scope}
				\end{scope}
			\end{tikzpicture}
			\label{eq:app_hamil_2}
		\end{equation}
		where each diagonal is two-periodic, and can be extended to describe nanoribbons of arbitrary width. To obtain the modified rhombic termination depicted in Figs.~\ref{fig:termination}(m)-(p), the first atom has to be flipped, moving the `dangling' bond to the other end. Therefore, $H^{(4)}_z(k)$ is similar to $H^{(3)}_z(k)$, but as before the first $3\times3$ subblock (marked blue in Eq.~\ref{eq:app_hamil_2}) is changed.
		\begin{equation}
			H^{(4)}_z(k)_{1:3, 1:3} =
			\begin{pmatrix}
				M + g(\phi, k) & t(1+e^{-ik}) & f(-\phi, -k) & \\
				t(1+e^{ik}) & -M +g(-\phi, k) & t  \\
				f(\phi, k) & t & M + g(\phi, k)
			\end{pmatrix}
		\end{equation}

		\section{Additional end terminations for armchair nanoribbons}\label{app:TerminationArmchair}
		For the sake of completeness, here we report the energy gap, Zak phase, and open spectra for two different end terminations of armchair nanoribbons. In Fig.~\ref{fig:at1}, an armchair nanoribbon is considered, whose unit cell is shifted by $a/2$ compared to the armchair nanoribbon in Fig.~\ref{fig:armchair}. The results are presented in complete analogy with Fig.~\ref{fig:armchair}, and again, no pinned end states are observed and the Zak phase does not quantize. 
		\begin{figure}[H]
			\centering
			\includegraphics[width=0.8\textwidth]{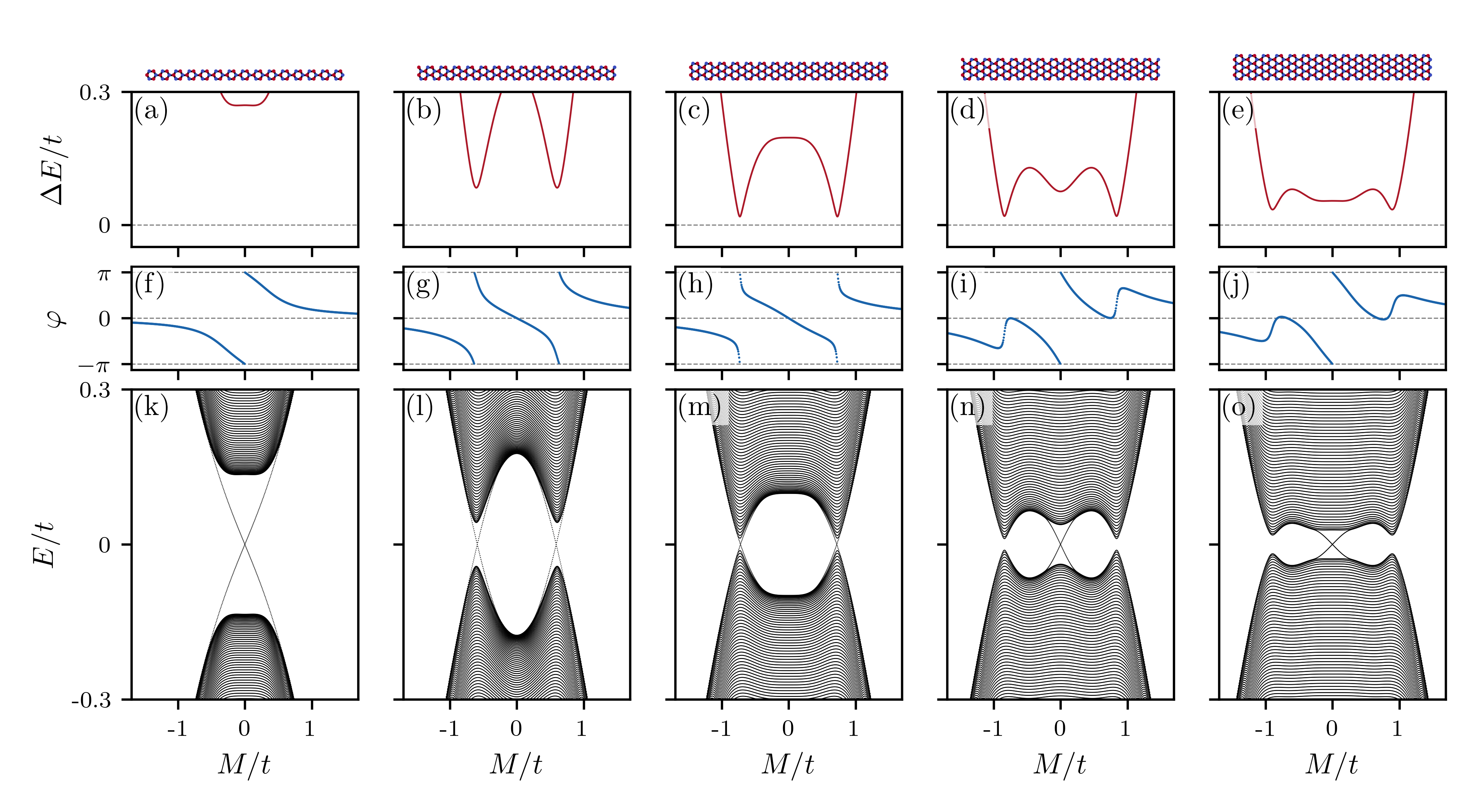}
			\caption{The bulk energy gap, Zak phase, and open spectrum for the half-unit-cell shifted armchair nanoribbon. The widths considered are: (a, f, k) 1-, (b, g, l) 1.5-, (c, h, m) 2-, (d, i, n) 2.5- and (e, j, o) 3-hexagon-wide nanoribbons.}
			\label{fig:at1}
		\end{figure}
		
		Similarly, an armchair nanoribbon with a sharp point end termination was considered. These results are presented in Fig.~\ref{fig:at2}. Here, we only consider odd widths to keep them symmetric. No quantization of the Zak phase occurs, but non-pinned end states are present.
		\begin{figure}[H]
			\centering
			\includegraphics[width=0.8\textwidth]{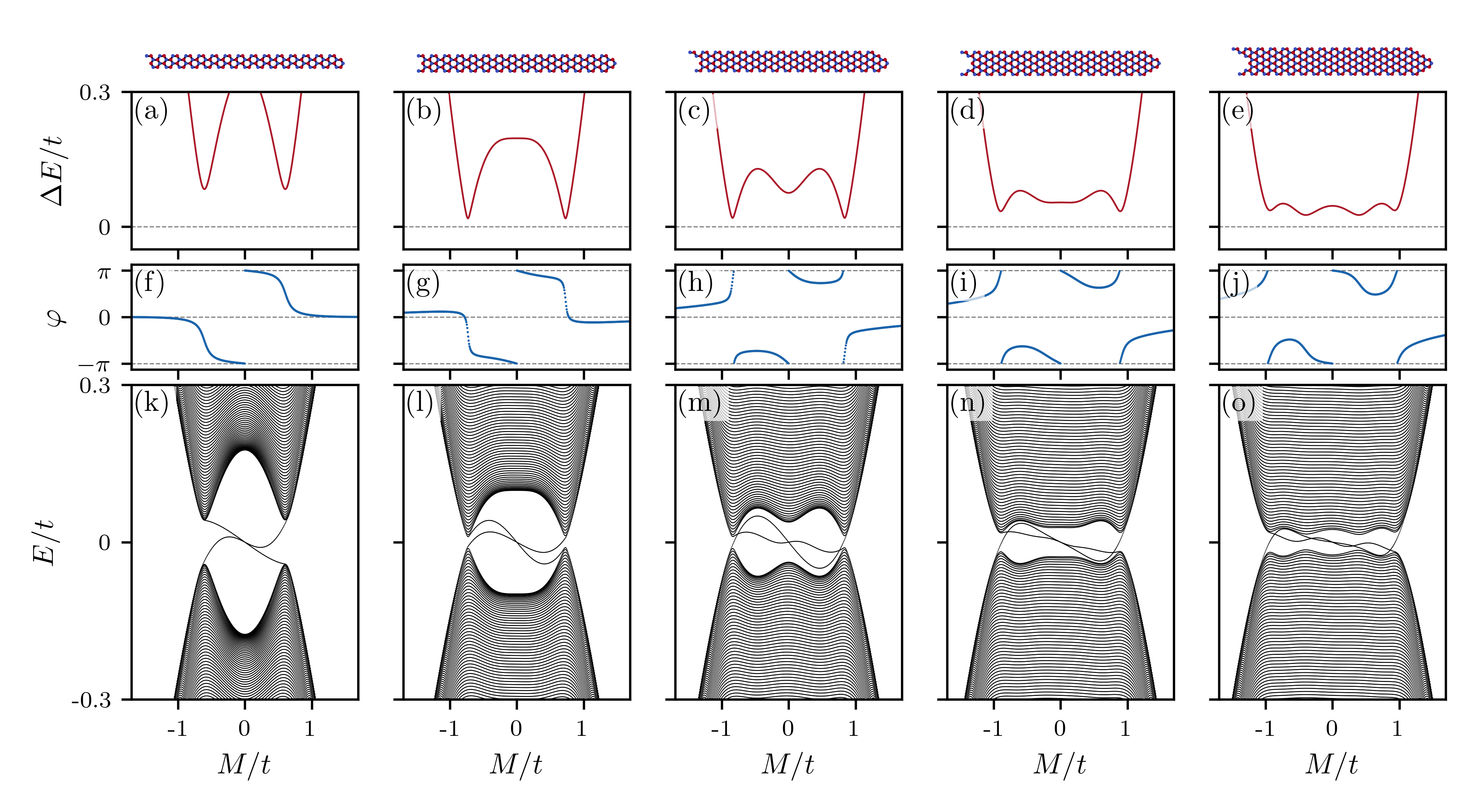}
			\caption{The bulk energy gap, Zak phase, and open spectrum for the sharply terminated armchair nanoribbon. The widths considered are: (a, f, k) 1-, (b, g, l) 1.5-, (c, h, m) 2-, (d, i, n) 2.5- and (e, j, o) 3-hexagon-wide nanoribbons.}
			\label{fig:at2}
		\end{figure}
		
	\end{widetext}
	
	\bibliography{Master}

@article{FinoHybrid,
  title = {Hybrid interacting quantum {Hall} thermal machine},
  author = {Finocchiaro, S. and Ferraro, D. and Sassetti, M. and Benenti, G.},
  journal = {Phys. Rev. B},
  volume = {111},
  issue = {20},
  pages = {205420},
  numpages = {13},
  year = {2025},
  month = {May},
  publisher = {American Physical Society},
  doi = {10.1103/PhysRevB.111.205420},
  url = {https://link.aps.org/doi/10.1103/PhysRevB.111.205420}
}

@article{LeeFloquet,
  title = {Floquet engineering of topological phase transitions in a quantum spin {H}all $\ensuremath{\alpha}\text{\ensuremath{-}}{T}_{3}$ system},
  author = {Lee, Kok Wai and Calderon, Mateo Jalen Andrew and Yu, Xiang-Long and Lee, Ching Hua and Ang, Yee Sin and Fu, Pei-Hao},
  journal = {Phys. Rev. B},
  volume = {111},
  issue = {4},
  pages = {045406},
  numpages = {12},
  year = {2025},
  month = {Jan},
  publisher = {American Physical Society},
  doi = {10.1103/PhysRevB.111.045406},
  url = {https://link.aps.org/doi/10.1103/PhysRevB.111.045406}
}

@article{LeeModified,
  title = {Interplay between {H}aldane and modified {H}aldane models in $\ensuremath{\alpha}\text{\ensuremath{-}}{T}_{3}$ lattice: Band structures, phase diagrams, and edge states},
  author = {Lee, Kok Wai and Fu, Pei-Hao and Ang, Yee Sin},
  journal = {Phys. Rev. B},
  volume = {109},
  issue = {23},
  pages = {235105},
  numpages = {11},
  year = {2024},
  month = {Jun},
  publisher = {American Physical Society},
  doi = {10.1103/PhysRevB.109.235105},
  url = {https://link.aps.org/doi/10.1103/PhysRevB.109.235105}
}

@article{haldane_model_1988,
	title = {Model for a {Quantum} {Hall} {Effect} without {Landau} {Levels}: {Condensed}-{Matter} {Realization} of the ``{Parity} {Anomaly}"},
	volume = {61},
	copyright = {http://link.aps.org/licenses/aps-default-license},
	issn = {0031-9007},
	shorttitle = {Model for a {Quantum} {Hall} {Effect} without {Landau} {Levels}},
	url = {https://link.aps.org/doi/10.1103/PhysRevLett.61.2015},
	doi = {10.1103/PhysRevLett.61.2015},
	language = {english},
	number = {18},
	urldate = {2024-05-21},
	journal = {Phys. Rev. Lett.},
	author = {Haldane, F. D. M.},
	month = oct,
	year = {1988},
	pages = {2015--2018},
}

@article{su_solitons_1979,
	title = {Solitons in {Polyacetylene}},
	volume = {42},
	copyright = {http://link.aps.org/licenses/aps-default-license},
	issn = {0031-9007},
	url = {https://link.aps.org/doi/10.1103/PhysRevLett.42.1698},
	doi = {10.1103/PhysRevLett.42.1698},
	language = {english},
	number = {25},
	urldate = {2025-09-02},
	journal = {Phys. Rev. Lett.},
	author = {Su, W. P. and Schrieffer, J. R. and Heeger, A. J.},
	month = jun,
	year = {1979},
	pages = {1698--1701},
}

@article{zak_berrys_1989,
	title = {Berry’s phase for energy bands in solids},
	volume = {62},
	copyright = {http://link.aps.org/licenses/aps-default-license},
	issn = {0031-9007},
	url = {https://link.aps.org/doi/10.1103/PhysRevLett.62.2747},
	doi = {10.1103/PhysRevLett.62.2747},
	number = {23},
	urldate = {2025-09-24},
	journal = {Phys. Rev. Lett.},
	author = {Zak, J.},
	month = jun,
	year = {1989},
	pages = {2747--2750},
}

@article{soluyanov_smooth_2012,
	title = {Smooth gauge for topological insulators},
	volume = {85},
	copyright = {http://link.aps.org/licenses/aps-default-license},
	issn = {1098-0121, 1550-235X},
	url = {https://link.aps.org/doi/10.1103/PhysRevB.85.115415},
	doi = {10.1103/PhysRevB.85.115415},
	language = {english},
	number = {11},
	urldate = {2025-09-24},
	journal = {Phys. Rev. B},
	author = {Soluyanov, Alexey A. and Vanderbilt, David},
	month = mar,
	year = {2012},
	pages = {115415},
}

@article{traverso_emerging_2024,
	title = {Emerging topological bound states in {Haldane} model zigzag nanoribbons},
	volume = {9},
	issn = {2397-4648},
	url = {https://www.nature.com/articles/s41535-023-00615-1},
	doi = {10.1038/s41535-023-00615-1},
	number = {1},
	urldate = {2025-09-24},
	journal = {npj Quantum Mater.},
	author = {Traverso, Simone and Sassetti, Maura and Traverso Ziani, Niccolò},
	month = jan,
	year = {2024},
	pages = {9},
}

@book{vanderbilt_berry_2018,
	edition = {1},
	title = {Berry {Phases} in {Electronic} {Structure} {Theory}: {Electric} {Polarization}, {Orbital} {Magnetization} and {Topological} {Insulators}},
	copyright = {https://www.cambridge.org/core/terms},
	isbn = {978-1-316-66220-5 978-1-107-15765-1},
	shorttitle = {Berry {Phases} in {Electronic} {Structure} {Theory}},
	url = {https://www.cambridge.org/core/product/identifier/9781316662205/type/book},
	urldate = {2025-09-24},
	publisher = {Cambridge University Press},
	author = {Vanderbilt, David},
	month = oct,
	year = {2018},
	doi = {10.1017/9781316662205},
}

@article{eek_fractality-induced_2025,
	title = {Fractality-{Induced} {Topology}},
	volume = {134},
	issn = {0031-9007, 1079-7114},
	url = {https://link.aps.org/doi/10.1103/jcrl-9dz6},
	doi = {10.1103/jcrl-9dz6},
	language = {english},
	number = {24},
	urldate = {2025-09-26},
	journal = {Phys. Rev. Lett.},
	author = {Eek, L. and Osseweijer, Z.F. and Morais Smith, C.},
	month = jun,
	year = {2025},
	pages = {246601},
}

@article{osseweijer_haldane_2024,
	title = {Haldane model on the {Sierpiński} gasket},
	volume = {110},
	issn = {2469-9950, 2469-9969},
	url = {https://link.aps.org/doi/10.1103/PhysRevB.110.245405},
	doi = {10.1103/PhysRevB.110.245405},
	language = {english},
	number = {24},
	urldate = {2025-09-26},
	journal = {Phys. Rev. B},
	author = {Osseweijer, Z. F. and Eek, L. and Moustaj, A. and Fremling, M. and Morais Smith, C.},
	month = dec,
	year = {2024},
	pages = {245405},
}

@article{kaneTopologicalOrderQuantum2005,
  title = {{Z$_2$} {{Topological Order}} and the {{Quantum Spin Hall Effect}}},
  author = {Kane, C. L. and Mele, E. J.},
  year = 2005,
  month = sep,
  journal = {Phys. Rev. Lett.},
  volume = {95},
  number = {14},
  pages = {146802},
  issn = {0031-9007, 1079-7114},
  doi = {10.1103/PhysRevLett.95.146802},
  urldate = {2024-05-21},
  copyright = {http://link.aps.org/licenses/aps-default-license}
}

@article{kaneQuantumSpinHall2005,
  title = {Quantum {{Spin Hall Effect}} in {{Graphene}}},
  author = {Kane, C. L. and Mele, E. J.},
  year = 2005,
  month = nov,
  journal = {Phys. Rev. Lett.},
  volume = {95},
  number = {22},
  pages = {226801},
  issn = {0031-9007, 1079-7114},
  doi = {10.1103/PhysRevLett.95.226801},
  urldate = {2024-05-21},
  copyright = {http://link.aps.org/licenses/aps-default-license}
}

@article{klitzingNewMethodHighAccuracy1980,
  title = {New {{Method}} for {{High-Accuracy Determination}} of the {{Fine-Structure Constant Based}} on {{Quantized Hall Resistance}}},
  author = {Klitzing, K. {v}. and Dorda, G. and Pepper, M.},
  year = 1980,
  month = aug,
  journal = {Phys. Rev. Lett.},
  volume = {45},
  number = {6},
  pages = {494--497},
  issn = {0031-9007},
  doi = {10.1103/PhysRevLett.45.494},
  urldate = {2024-05-21},
  copyright = {http://link.aps.org/licenses/aps-default-license}
}

@article{term1,
  title = {Stable hydrogenated graphene edge types: Normal and reconstructed {Klein} edges},
  author = {Wagner, Philipp and Ivanovskaya, Viktoria V. and Melle-Franco, Manuel and Humbert, Bernard and Adjizian, Jean-Joseph and Briddon, Patrick R. and Ewels, Christopher P.},
  journal = {Phys. Rev. B},
  volume = {88},
  issue = {9},
  pages = {094106},
  numpages = {6},
  year = {2013},
  month = {Sep},
  publisher = {American Physical Society},
  doi = {10.1103/PhysRevB.88.094106},
  url = {https://link.aps.org/doi/10.1103/PhysRevB.88.094106}
}

@article{term2,
    author       = {Lou, S. and Lyu, B. and Zhou, X. and Shen, P. and Shi, Z.},
    title        = {Graphene nanoribbons: current status, challenges and opportunities},
    journal      = {Quantum Front},
    year         = {2024},
    volume       = {3},
    number       = {3},
    url          = {https://doi.org/10.1007/s44214-024-00050-8}
}

@article{Moustajpump,
    AUTHOR = {Moustaj, Anouar and Krebbekx, Julius and Morais Smith, Cristiane},
    TITLE = {Anomalous Polarization in One-Dimensional Aperiodic Insulators},
    JOURNAL = {Condens. Matter},
    VOLUME = {10},
    YEAR = {2025},
    NUMBER = {1},
    pages = {3},
    URL = {https://www.mdpi.com/2410-3896/10/1/3},
    ISSN = {2410-3896},
}

@article{Klaassen2025GermaneneTI,
  author       = {Klaassen, Dennis J. and Eek, Lumen and Rudenko, Alexander N. and van 't Westende, Esra D. and Castenmiller, Carolien and Zhang, Zhiguo and de Boeij, Paul L. and van Houselt, Arie and Ezawa, Motohiko and Zandvliet, Harold J. W. and Morais Smith, Cristiane and Bampoulis, Pantelis},
  title        = {Realization of a one-dimensional topological insulator in ultrathin germanene nanoribbons},
  journal      = {Nat Commun},
  volume       = {16},
  pages        = {2059},
  year         = {2025},
}

@article{Eek2025Electric,
  title = {Electric-Field Control of Zero-Dimensional Topological States in Ultranarrow Germanene Nanoribbons},
  author = {Eek, Lumen and van 't Westende, Esra D. and Klaassen, Dennis J. and Zandvliet, Harold J. W. and Bampoulis, Pantelis and Smith, Cristiane Morais},
  journal = {Phys. Rev. Lett.},
  volume = {135},
  issue = {20},
  pages = {206601},
  numpages = {8},
  year = {2025},
  month = {Nov},
  publisher = {American Physical Society},
  doi = {10.1103/jx2x-fb5b},
  url = {https://link.aps.org/doi/10.1103/jx2x-fb5b}
}

@article{Schindler,
	author = {Frank Schindler and Ashley M. Cook and Maia G. Vergniory and Zhijun Wang and Stuart S. P. Parkin and B. Andrei Bernevig and Titus Neupert},
	doi = {10.1126/sciadv.aat0346},
	journal = {Sci. Adv.},
	number = {6},
	pages = {eaat0346},
	title = {Higher-order topological insulators},
	volume = {4},
	year = {2018},
}

@article{Cutting,
  title = {Appearance of hinge states in second-order topological insulators via the cutting procedure},
  author = {Tanaka, Yutaro and Takahashi, Ryo and Murakami, Shuichi},
  journal = {Phys. Rev. B},
  volume = {101},
  issue = {11},
  pages = {115120},
  numpages = {16},
  year = {2020},
  month = {Mar},
  publisher = {American Physical Society},
  doi = {10.1103/PhysRevB.101.115120},
  url = {https://link.aps.org/doi/10.1103/PhysRevB.101.115120}
}

@article{TuningHOTI,
  title = {Tuning three-dimensional higher-order topological insulators by surface state hybridization},
  author = {Lin, Hao-Jie and Sun, Hai-Peng and Liu, Tianyu and Zhao, Peng-Lu},
  journal = {Phys. Rev. B},
  volume = {108},
  issue = {16},
  pages = {165427},
  numpages = {9},
  year = {2023},
  month = {Oct},
  publisher = {American Physical Society},
  doi = {10.1103/PhysRevB.108.165427},
  url = {https://link.aps.org/doi/10.1103/PhysRevB.108.165427}
}

@article{MoesBi2Se3,
	author = {Moes, Jesper R. and Vliem, Jara F. and de Melo, Pedro M. M. C. and Wigmans, Thomas C. and Botello-M{\'e}ndez, Andr{\'e}s R. and Mendes, Rafael G. and van Brenk, Ella F. and Swart, Ingmar and Maisel Licer{\'a}n, Lucas and Stoof, Henk T. C. and Delerue, Christophe and Zanolli, Zeila and Vanmaekelbergh, Daniel},
	journal = {Nano Lett.},
	number = {17},
	pages = {5110-5116},
	title = {Characterization of the {Edge States} in {Colloidal} {Bi$_2$Se$_3$} {Platelets}},
	volume = {24},
	year = {2024},
}

@article{Zhang2010Bi2Se3Crossover,
  author       = {Zhang, Yi and He, Ke and Chang, Cui-Zu and Song, Can-Li and Wang, Li-Li and Chen, Xi and Jia, Jin-Feng and Fang, Zhong and Dai, Xi and Shan, Wen-Yu and Shen, Shun-Qing and Niu, Qian and Qi, Xiao-Liang and Zhang, Shou-Cheng and Ma, Xu-Cun and Xue, Qi-Kun},
  title        = {Crossover of the Three-Dimensional Topological Insulator {Bi$_2$Se$_3$} to the Two-Dimensional Limit},
  journal      = {Nat. Phys.},
  volume       = {6},
  pages        = {584--588},
  year         = {2010},
  doi          = {10.1038/nphys1689},
  url          = {https://doi.org/10.1038/nphys1689}
}

@article{LiuOscillatory,
  title = {Oscillatory crossover from two-dimensional to three-dimensional topological insulators},
  author = {Liu, Chao-Xing and Zhang, HaiJun and Yan, Binghai and Qi, Xiao-Liang and Frauenheim, Thomas and Dai, Xi and Fang, Zhong and Zhang, Shou-Cheng},
  journal = {Phys. Rev. B},
  volume = {81},
  issue = {4},
  pages = {041307},
  numpages = {4},
  year = {2010},
  month = {Jan},
  publisher = {American Physical Society},
  doi = {10.1103/PhysRevB.81.041307},
  url = {https://link.aps.org/doi/10.1103/PhysRevB.81.041307}
}

@article{bismuthene,
	author = {F. Reis and G. Li and L. Dudy and M. Bauernfeind and S. Glass and W. Hanke and R. Thomale and J. Sch{\"a}fer and R. Claessen},
	journal = {Science},
	number = {6348},
	pages = {287-290},
	title = {Bismuthene on a {S}i{C} substrate: A candidate for a high-temperature quantum spin {H}all material},
	volume = {357},
	year = {2017},
}

@article{Tang2017WTe2QSH,
  author       = {Tang, Shujie and Zhang, Chaofan and Wong, Dillon and Pedramrazi, Zahra and Tsai, Hsin-Zon and Jia, Chunjing and Moritz, Brian and Claassen, Martin and Ryu, Hyejin and Kahn, Salman and Jiang, Juan and Yan, Hao and Hashimoto, Makoto and Lu, Donghui and Moore, Robert G. and Hwang, Chan-Cuk and Hwang, Choongyu and Hussain, Zahid and Chen, Yulin and Ugeda, Miguel M. and Liu, Zhi and Xie, Xiaoming and Devereaux, Thomas P. and Crommie, Michael F. and Mo, Sung-Kwan and Shen, Zhi-Xun},
  title        = {Quantum Spin {Hall} State in Monolayer 1{T}-{W}{T}e$_2$},
  journal      = {Nat. Phys.},
  volume       = {13},
  pages        = {683--687},
  year         = {2017},
  doi          = {10.1038/nphys4174},
  url          = {https://doi.org/10.1038/nphys4174}
}

@article{stanene,
  title = {Stable two-dimensional dumbbell stanene: A quantum spin {Hall} insulator},
  author = {Tang, Peizhe and Chen, Pengcheng and Cao, Wendong and Huang, Huaqing and Cahangirov, Seymur and Xian, Lede and Xu, Yong and Zhang, Shou-Cheng and Duan, Wenhui and Rubio, Angel},
  journal = {Phys. Rev. B},
  volume = {90},
  issue = {12},
  pages = {121408},
  numpages = {6},
  year = {2014},
  month = {Sep},
  publisher = {American Physical Society},
  doi = {10.1103/PhysRevB.90.121408},
  url = {https://link.aps.org/doi/10.1103/PhysRevB.90.121408}
}

@article{Canyellas2024BismuthFractal,
  author       = {Canyellas, R. and Liu, Chen and Arouca, R. and Eek, L. and Wang, Guanyong and Yin, Yin and Guan, Dandan and Li, Yaoyi and Wang, Shiyong and Zheng, Hao and Liu, Canhua and Jia, Jinfeng and Morais Smith, C.},
  title        = {Topological edge and corner states in bismuth fractal nanostructures},
  journal      = {Nat. Phys.},
  volume       = {20},
  pages        = {1421--1428},
  year         = {2024},
  doi          = {10.1038/s41567-024-01547-3},
  url          = {https://doi.org/10.1038/s41567-024-01547-3}
}

@article{Tenfoldway,
  title = {Nonstandard symmetry classes in mesoscopic normal-superconducting hybrid structures},
  author = {Altland, Alexander and Zirnbauer, Martin R.},
  journal = {Phys. Rev. B},
  volume = {55},
  issue = {2},
  pages = {1142--1161},
  numpages = {0},
  year = {1997},
  month = {Jan},
  publisher = {American Physical Society},
  doi = {10.1103/PhysRevB.55.1142},
  url = {https://link.aps.org/doi/10.1103/PhysRevB.55.1142}
}

@article{winding,
  title = {Classification of topological quantum matter with symmetries},
  author = {Chiu, Ching-Kai and Teo, Jeffrey C. Y. and Schnyder, Andreas P. and Ryu, Shinsei},
  journal = {Rev. Mod. Phys.},
  volume = {88},
  issue = {3},
  pages = {035005},
  numpages = {63},
  year = {2016},
  month = {Aug},
  publisher = {American Physical Society},
  doi = {10.1103/RevModPhys.88.035005},
  url = {https://link.aps.org/doi/10.1103/RevModPhys.88.035005}
}

@article{parto_edge-mode_2018,
	title = {Edge-{Mode} {Lasing} in {1D} {Topological} {Active} {Arrays}},
	volume = {120},
	issn = {0031-9007, 1079-7114},
	doi = {10.1103/PhysRevLett.120.113901},
	number = {11},
	urldate = {2026-02-22},
	journal = {Phys. Rev. Lett.},
	author = {Parto, Midya and Wittek, Steffen and Hodaei, Hossein and Harari, Gal and Bandres, Miguel A. and Ren, Jinhan and Rechtsman, Mikael C. and Segev, Mordechai and Christodoulides, Demetrios N. and Khajavikhan, Mercedeh},
	month = mar,
	year = {2018},
	pages = {113901},
}

@article{zhao_topological_2018,
	title = {Topological hybrid silicon microlasers},
	volume = {9},
	issn = {2041-1723},
	doi = {10.1038/s41467-018-03434-2},
	number = {1},
	journal = {Nat Commun},
	author = {Zhao, Han and Miao, Pei and Teimourpour, Mohammad H. and Malzard, Simon and El-Ganainy, Ramy and Schomerus, Henning and Feng, Liang},
	month = mar,
	year = {2018},
	pages = {981},
}

@article{henriques_topological_2020,
	title = {Topological photonic {Tamm} states and the {Su}-{Schrieffer}-{Heeger} model},
	volume = {101},
	issn = {2469-9926, 2469-9934},
	doi = {10.1103/PhysRevA.101.043811},
	number = {4},
	journal = {Phys. Rev. A},
	author = {Henriques, J. C. G. and Rappoport, T. G. and Bludov, Y. V. and Vasilevskiy, M. I. and Peres, N. M. R.},
	month = apr,
	year = {2020},
	pages = {043811},
}

@article{PhysRevResearch.6.023140,
  title = {Breaking and resurgence of symmetry in the non-{H}ermitian {S}u-{S}chrieffer-{H}eeger model in photonic waveguides},
  author = {Slootman, E. and Cherifi, W. and Eek, L. and Arouca, R. and Bergholtz, E. J. and Bourennane, M. and Smith, C. Morais},
  journal = {Phys. Rev. Res.},
  volume = {6},
  issue = {2},
  pages = {023140},
  numpages = {23},
  year = {2024},
  month = {May},
  publisher = {American Physical Society},
  doi = {10.1103/PhysRevResearch.6.023140},
}

@article{PlateroMonomode,
  title = {Optimizing edge-state transfer in a {Su-Schrieffer-Heeger} chain via hybrid analog-digital strategies},
  author = {Romero, Sebasti\'an V. and Chen, Xi and Platero, Gloria and Ban, Yue},
  journal = {Phys. Rev. Appl.},
  volume = {21},
  issue = {3},
  pages = {034033},
  numpages = {19},
  year = {2024},
  month = {Mar},
  publisher = {American Physical Society},
  doi = {10.1103/PhysRevApplied.21.034033},
  url = {https://link.aps.org/doi/10.1103/PhysRevApplied.21.034033}
}

@article{fu_topological_2011,
	title = {Topological {crystalline} {insulators}},
	volume = {106},
	doi = {10.1103/PhysRevLett.106.106802},
	number = {10},
	urldate = {2026-02-21},
	journal = {Phys. Rev. Lett.},
	author = {Fu, Liang},
	month = mar,
	year = {2011},
	pages = {106802},
}

\end{document}